\documentclass[11pt,thmsb]{article}
\usepackage{amsfonts}
\usepackage{amssymb}
\usepackage{amsmath}
\usepackage{harvard}
\usepackage{setspace}
\usepackage{dirtytalk}
\usepackage{graphicx}
\usepackage{multirow}
\usepackage{rotating}
\usepackage{lscape}
\usepackage{caption}
\usepackage{subcaption}
\usepackage[affil-it]{authblk}
\usepackage[flushleft]{threeparttable}
\advance \topmargin by -\headheight
\advance \topmargin by -\headsep
\evensidemargin \oddsidemargin
\let\tilde=\widetilde

\input{tcilatex}
\makeatletter
\def\@biblabel#1{\hspace*{-\labelsep}}
\makeatother

\title{Pairs Trading with Nonlinear and Non-Gaussian State Space Models\thanks{I am grateful to Zhongjun Qu, Hiroaki Kaido, Jean-Jacques
Forneron and seminal participants at the Boston University Economics Department.}}

\author{Guang Zhang\thanks{%
Email: gzhang46@bu.edu. }} 
\affil{Department of Economics, Boston University, Boston, MA, 02215}
\date{\today }

\begin{document}
\maketitle

\begin{abstract}
\baselineskip=17pt 
%In this paper, we carry out a study of pairs trading based on a nonlinear
%and non-Gaussian state space modeling. We model the spread between
%the prices of two assets as an unobservable state variable and assume
%it follows a mean reverting process. This new model allows for (1)
%non-Gaussianity and heteroskedasticity of the noises effect the spread;
%(2) nonlinear mean-reversion of the spread. We then use filtered spread
%as the trading indicator for statistical arbitrage. To improve the
%performance, we also propose a new trading strategy and present a
%simulation based approach for the selection of optimal trading rules.
%We implement this trading approach to two examples: PEP vs KO and
%EWT vs EWH, and our empirical results show the new approach for pairs trading proposed by this paper can achieve 21.86\%
%annualized return for the PEP/KO pair and 31.84\% annualized return
%for the EWT/EWH pair. We also test our approach on all possible pairs among biggest 5 US banks and smallest 5 US banks listed on NYSE. We compare the performance of the approach proposed in this paper with that of existing popular approach on pairs trading, for both of the in-sample period and out-of-sample period, and find that our approach can significantly improve the return and Sharpe ratio of pairs trading.
This paper studies pairs trading using a nonlinear and non-Gaussian state space model
framework. We model the spread between the prices of two assets as an unobservable state
variable, and assume that it follows a mean reverting process. This new model has two distinctive
features: (1) The innovations to the spread is non-Gaussianity and heteroskedastic. (2) The mean
reversion of the spread is nonlinear. We show how to use the filtered spread as the trading indicator
to carry out statistical arbitrage. We also propose a new trading strategy and present a Monte
Carlo based approach to select the optimal trading rule. As the first empirical application, we apply
the new model and the new trading strategy to two examples: PEP vs KO and EWT vs EWH. The
results show that the new approach can achieve 21.86\% annualized return for the PEP/KO pair
and 31.84\% annualized return for the EWT/EWH pair. As the second empirical application, we
consider all the possible pairs among the largest and the smallest five US banks listed on the NYSE.
For these pairs, we compare the performance of the proposed approach with that of the existing
popular approaches, both in-sample and out-of-sample. Interestingly, we find that our approach
can significantly improve the return and the Sharpe ratio in almost all the cases considered.


\end{abstract}
\thispagestyle{empty}\setcounter{page}{0}\baselineskip=18pt\newpage
\section{Introduction}

In early 1980s, a group of physicists, mathematicians and computer
scientists, leaded by quantitative analyst Nunzio Tartaglia, tried
to use a sophisticated statistical approach to find the opportunities
of arbitrage trading (Gatev et al. 2006). Tartaglia's strategy, later
coined pairs trading, is to find a pair of two stocks whose prices
have moved similarly historically, and make profit by applying the
simple contrarian principles. Since then, pairs trading has become
a popular short-term arbitrage strategy used by hedge funds and is often
considered as the ``ancestor'' of statistical arbitrage.

Pairs trading works by constructing a self financing portfolio with
a long position in one security and a short position in the other.
Given that the two securities have moved together historically, when a
temporary anomaly happens, one security would be overvalued than the other relative
to the long-term equilibrium. Then, an investor may be able to make money
by selling the overvalued security, buying the undervalued security,
and clearing the exposure when the two securities settle back to their
long-term equilibrium. Because the effect from movement of the market
is hedged by this self financing portfolio, pairs trading is
market-neutral.

The methods for pairs trading can be broadly divided into nonparametric and parametric methods. In particular, Gatev et al. (2006) propose a nonparametric distance based
approach in determining the securities for constructing the pairs. They choose a pair by finding the securities that minimized
the sum of squared deviations between the two normalized prices. They
argue this approach ``best approximates the description of how traders
themselves choose pairs''. They find that average annualized excess returns
reach 11\% for the top pairs portfolios using CRSP daily data from
1962 to 2002. Other Nonparametric methods on pairs trading can also be found in Bogomolov (2013) among others. Overall, the nonparametric distance based approach provides a simple and general method of selecting ``good'' pairs; however, as pointed out by Krauss (2016) and others, this selection metric is prone to pick up pairs with small variance of the spread, and therefore limits the profitability of pairs trading.

In contrast, the parametric approach tries to capture the mean-reverting  characteristic of the spread using a parametric model. For example, Elliott et al. (2005) propose a mean-reverting
Gaussian Markov chain model for the spread which is observed in Gaussian
noise. See Vidyamurthy (2004), Cummins and Bucca (2012), Tourin and Yan
(2013), Moura et al. (2016), Stübinger and Endres (2018), Clegg and Krauss (2018), Elliott and Bradrania (2018), Bai and Wu (2018) for other parametric methods on pairs trading. Overall, the parametric approach provides tractable methods for the analysis of pairs trading; however, most of the existing parametric models are too simple to be capable of capturing the dynamics of asset price, which substantially limits the returns from pairs trading.

Compared with the existing methods on pairs trading, the proposed approach has the following features: (1) It is based on a nonlinear and non-Gaussian state space model. This modelling can capture several stylized features of financial asset prices, including heavy-tailedness, heteroskedasticity, volatility clustering and nonlinear dependence. (2) The trading strategy is different from the existing ones. It utilizes the features of the model such as heteroskedasticity and volatility clustering, and it can potentially achieve significantly higher returns and Sharpe ratios. (3) The optimal trading rules is also different from the existing ones. Although this rule has no analytic solution, we show that it can be computed effectively using simulations. Finally, the optimal trading rule can adapt to various objectives, such as a high cumulative return, Sharpe ratio, or Calmar ratio.

We apply our approach to two pairs: PEP vs KO and EWT vs EWH. We we find that our approach achieves an annualized return of 0.2186 and Sharpe ratio of 2.9518 on the PEP/KO pair and an annualized return of 0.3184 and Sharpe ratio of 3.8892 on the EWT/EWH pair. In comparison, a conventional approach applied to the same pairs can only achieve an annualized return of 0.1311 and Sharpe ratio of 1.1003 for the PEP/KO pair and an annualized return of 0.1480 and Sharpe ratio of 1.1277 for the EWT/EWH pair. Next, we test our approach using all the possible pairs among the largest 5 banks and the smallest 5 banks listed in NYSE. We find significant improvements over the conventional approach for almost all the pairs. We also find that the pairs between small banks produce higher return than the pairs between large banks. This is likely because the spread between small banks are more volatile, providing more opportunities for active trading.

The main contributions of this paper can be summarized as follows. On the theory side, we propose a complete set of tools for pairs trading that include a model for the dynamics of the spread, a new trading strategy and a Monte Carlo method for determining the optimal trading rule. On the empirical side, we apply our approach to various pairs in practice. The results show that the new approach can achieve significant improvements on the performance of pairs trading.

The remainder of this paper is organized as follows. In Section 2, we propose
a new model for pairs trading. In Section 3, we propose a new trading
strategy based on the mean-reverting property of spread, and compare it with conventional trading strategies using simulations. In Section 4, we implement the proposed approach to actual data, and in Section 5 we conclude the paper. 

\section{A New Model for Pairs Trading}

We propose the following nonlinear
and non-Gaussian state space model for pairs trading:
\begin{eqnarray}
P_{A,t} & = & \phi+\gamma P_{B,t}+x_{t}+\varepsilon_{t}\label{eq:pt1}\\
x_{t+1} & = & f\left(x_{t};\theta\right)+g\left(x_{t};\theta\right)*\eta_{t}\label{eq:pt2}
\end{eqnarray}
where $P_{A}$ is the price of security $A$, $P_{B}$ is the price
of security $B$, $\gamma$ is the hedge ratio between two securities,
and $x$ is the true spread between $P_{A}$ and $P_{B}$. We assume
$x$ follow a mean-reverting process as in (\ref{eq:pt2}), $\varepsilon_{t}\sim N\left(0,\sigma_{\varepsilon}^{2}\right)$
and $\eta_{t}\sim p\left(\eta_{t};\theta\right)$ which could be non-Gaussian.
Popular choices for $f$, $g$ and $p$ could be the followings. Our framework applies to all of them.
\begin{itemize}
\item Linear mean-reverting (Ornstein--Uhlenbeck process): $f\left(x_{t};\theta\right)=\theta_{1}+\theta_{2}x_{t}$
\item Nonlinear mean-reverting model: $f\left(x_{t};\theta\right)=\theta_{1}+\theta_{2}x_{t}+\theta_{3}x_{t}^{2}$
\item Ait-Sahalia's nonlinear mean-reverting model (Ait-Sahalia, 1996):
$f\left(x_{t};\theta\right)=\theta_{1}+\theta_{2}x_{t}^{-1}+\theta_{3}x_{t}+\theta_{4}x_{t}^{2}$
\item Homoskedasticity model: $g\left(x_{t};\theta\right)=1$
\item ARCH$(m)$ model: $g\left(x_{t};\theta\right)=\sqrt{\theta_{0}+\sum_{i=1}^{m}\theta_{i}x_{t-i}^{2}}$
\item APARCH$(m,\delta)$ model: $g\left(x_{t};\theta\right)=\left(\theta_{0}+\sum_{i=1}^{m}\theta_{i}\mid x_{t-i}\mid^{\delta}\right)^{\frac{1}{\delta}}$
\item Gaussian distributed noise: $p\left(\eta;\mu,\sigma\right)=\frac{1}{\sqrt{2\pi}\sigma}\exp\left(-\frac{\left(\mu-\eta\right)^{2}}{2\sigma^{2}}\right)$
\item Student's $t$ distributed noise: $p\left(\eta;\nu\right)=\frac{\Gamma\left(\frac{\nu+1}{2}\right)}{\sqrt{\nu\pi}\Gamma\left(\frac{\nu}{2}\right)}\left(1+\frac{\eta^{2}}{\nu}\right)^{-\frac{\nu+1}{2}}$
\item Generalized error distributed noise: $p\left(\eta;\alpha,\beta,\mu\right)=\frac{\beta}{2\alpha\Gamma\left(\frac{1}{\beta}\right)}\exp\left(-\left(\mid\eta-\mu\mid/\alpha\right)^{\beta}\right)$
\end{itemize}
In model (\ref{eq:pt1})-(\ref{eq:pt2}), we consider $x$ as the unobservable
true spread between security $A$ and $B$, which follows a mean-reverting
process. $P_{A}$ is the observation and $P_{B}$ is the control variable.
Since $\phi$ and $\theta_{1}$ in the $f$ function can not be identified simultaneously, we let $\phi=0$ and denote $\psi=\left(\gamma,\theta,\sigma_{\varepsilon}\right)$
as the parameter of the model (\ref{eq:pt1})-(\ref{eq:pt2}). $\psi$
is going to determined based on data set $\left\{ P_{A,t},P_{B,t}\right\} _{t=0}^{T}$

Our new model has three advantages compared with existing models for
pairs trading, such as Elliott et al. (2005) and Moura et al. (2016).
First, since $\eta$ can be non-Gaussian, $x$ can follow a non-Gaussian process. By allowing for this non-Gaussianity in $\eta$, the model
can capture the distributional deviation from Gaussianity and reproduce
heavy-tailed returns.

Second, the model captures heteroskedasticity
in financial data. A well-known feature of financial time-series is
volatility clustering: \say{large changes tend to be followed
by large changes, of either sign, and small changes tend to be followed
by small changes} (Mandelbrot, 1963). This feature was
documented later in Ding, Granger and Engle (1993), and Ding and Granger
(1996) among others. In model (\ref{eq:pt2}), the volatility persistence
is represented by ARCH-style modeling. Details about the application
of ARCH model in finance can be found in Bollerslev, Chou and Kroner
(1992).

Third, in order to characterize the nonlinear dependence in financial data, we allow $f$ to be nonlinear. Scheinkman and LeBaron (1989) find evidence
that indicates the presence of nonlinear dependence in weekly returns
on the CRSP value-weighted index. Ait-Sahalia (1996) finds
nonlinearity in the drift function of interest rate and concludes
that ``the principal source of rejection of existing (linear drift)
models is the strong nonlinearity of the drift''. We keep the functional form of $f$ flexible and, as a result, we can capture the nonlinear dependence in financial data.

\section{A New Approach to Pairs Trading}
In this section, we discuss the trading strategies and trading rules for pairs trading. In this paper, a trading strategy is the method of buying and selling of assets in markets based on the estimation of the unobservable spread. A trading rule is the predefined values to generate the trading signal for a specific trading strategy with an investing objective. To implement a strategy and rule on pairs trading, we need the following quantities: (i) parameter estimates for the model (\ref{eq:pt1})-(\ref{eq:pt2}), (ii) an estimate of the spread, and (iii) choice of a specific strategy and the optimal trading rule, and we discuss these aspects in this section. More specifically, in Section 3.1, we present an algorithm on the filtering of the unobervable spread and parameter estimation. In Section 3.2, We will discuss two benchmark trading strategies. In Section 3.3, we will present and compare three popular trading rules associated with the benchmark trading strategies. In Section 3.4, we propose a new trading strategy. In this new trading strategy, we change the way we open or close a trade, and we will discuss the benefit of this new strategy compared with the benchmark strategies. Since the existing trading rule cannot be simply applied to the model (\ref{eq:pt1})-(\ref{eq:pt2}), we propose a new approach to calculate the optimal trading rule based on the simulation of the spread. The detail of this simulation based method is in Section 3.5. In Section 3.6, we summarize the procedure of pairs trading. This procedure can be applied to pairs trading with all of the trading strategies and trading rules discussed in this paper.

\subsection{Algorithm for Filtering and Parameter Estimation}
For a specification of model (\ref{eq:pt1})-(\ref{eq:pt2}), we run the
following algorithm of Quasi Monte Carlo Kalman filter for nonlinear and non-Gaussian state space models to estimate the unobservable spread and unknown parameters
in the model, based on the observations $\left\{ P_{A,t},P_{B,t}\right\} _{t=0}^{T}$.
Suppose the initial spread $x_{0}$ follows $N\left(\mu,\varSigma\right)$ for any reasonable choices of $\mu$ and $\varSigma$.
\begin{itemize}
\item Step 1: For non-Gaussian density $p\left(\eta_{t}\right),$we use
Gaussian mixture density to approximate its pdf and denote the approximation
as $\tilde{p}\left(\eta_{t}\right)=\sum_{i=1}^{m}\alpha_{i}\phi\left(\eta_{t}-\mathrm{a}_{i},\mathrm{P}_{i}\right),\sum_{i=1}^{m}\alpha_{i}=1$
where $\phi$ is the Gaussian pdf defined by
\[
\phi\left(v,\Sigma\right)=\frac{1}{(2\pi)^{1/2}|\Sigma|^{1/2}}\exp\left(-\frac{1}{2}v^{T}\Sigma^{-1}v\right).
\]
To get this approximation, we determine the values of $\left\{ \alpha_{i},\mathrm{a}_{i},\mathrm{P}_{i}\right\} _{i=1}^{m}$
by minimizing the relative entropy between the true density $p\left(\eta_{t}\right)$
and its approximation $\tilde{p}\left(\eta_{t}\right)$. The relative
entropy is defined by 
\[
\mathcal{H}\left(p_{\mathbf{}}|\tilde{p}_{\mathbf{}}\right)=\int\left(\log\frac{p_{\mathbf{}}\left(\eta\right)}{\tilde{p}_{\mathbf{}}\left(\eta\right)}\right)\times p_{\mathbf{}}\left(\eta\right)d\eta.
\]
If $\eta_{t}$ is Gaussian, then this step can be dropped.
\item Step 2: Generate a Box-Muller transformed Halton sequence $\small\{ x_{t}^{\left(g\right)}\small\} _{g=1}^{G}$
with sequence size $G$ from $\phi\left(x_{t}-\mathrm{b}_{ts},\mathrm{P}_{ts}\right)$.
Compute and store 
\[
\mathrm{Q}_{t+1i}=\frac{1}{G}\sum_{g=1}^{G}\left(f\left(x_{t}^{\left(g\right)}\right)-\mathrm{c}_{t+1i}\right)^{2}+\left(g\left(x_{t}^{\left(g\right)}\right)\right)^{2}*\mathrm{P}_{k},
\]
and 
\[
\mathrm{c}_{t+1i}=\frac{1}{G}\sum_{g=1}^{G}f\left(x_{t}^{\left(g\right)}\right)+g\left(x_{t}^{\left(g\right)}\right)*\mathrm{a}_{k}.
\]
When $t=0$, $\small\{ x_{0}^{\left(g\right)}\small\} _{g=1}^{G}$
is sampled from $N\left(\mu,\varSigma\right)$.
\item Step 3: Repeat Step 2 for $s=1,2,...,J_{t+1}$, $J_{t+1}=m^{t}$,
and $k=1,\ldots m,$ and store $\mathrm{c}_{t+1i}$ and $\mathrm{Q}_{t+1i}$
for $i=1,2,...,I_{t+1},\:I_{t+1}=J_{t+1}*m=m^{t+1}$.
\item Step 4: Based on the results from Step 3, generate a Box-Muller transformed
Halton sequences $\small\{ x_{t+1i}^{\left(g\right)}\small\} _{g=1}^{G}$
from $\phi\left(x_{t+1}-\mathrm{c}_{t+1i},\mathrm{Q}_{t+1i}\right)$
for $i=1,2,...,I_{t+1},\:I_{t+1}=m^{t+1}$. Then generate $P_{A,t+1i}^{\left(g\right)}=x_{t+1i}^{\left(g\right)}+\gamma*P_{B,t+1}$.
Compute and store the followings 
\[
\bar{P}_{A,t+1i}=\frac{1}{G}\sum_{g=1}^{G}P_{A,t+1i}^{\left(g\right)},
\]
\[
\mathrm{V}_{t+1i}=\frac{1}{G}\sum_{g=1}^{G}\left(P_{A,t+1i}^{\left(g\right)}-\bar{P}_{A,t+1i}\right)^{2}
+\sigma_{\varepsilon}^{2},
\]
\[
\mathrm{S}_{t+1i}=\frac{1}{G}\sum_{g=1}^{G}\left(x_{t+1i}^{\left(g\right)}-\mathrm{c}_{t+1i}\right)\left(P_{A,t+1i}^{\left(g\right)}-\bar{P}_{A,t+1i}\right).
\]
\item Step 5: Compute $\mathrm{K}_{t+1i}=\mathrm{S}_{t+1i}\mathrm{V}_{t+1i}^{-1}$,
$\mathrm{P}_{t+1i}=\mathrm{Q}_{t+1i}-\mathrm{K}_{t+1i}^{2}\mathrm{V}_{t+1i}$,
and $\mathrm{b}_{t+1i}=\mathrm{c}_{t+1i}+\mathrm{K}_{t+1i}\left(P_{A,t+1}-\bar{P}_{A,t+1i}\right)$.
\item Step 6: Repeat Step 4-5 for $i=1,2,...,I_{t+1},\:I_{t+1}=m^{t+1}$.
Compute and store $\bar{x}_{t+1}$ and $\bar{P}_{t+1}$ where $\bar{x}_{t+1}=\sum_{i=1}^{I_{t+1}}\beta_{t+1i}\mathrm{b}_{t+1i}$
, and 
\[
\bar{P}_{t+1}=\sum_{i=1}^{I_{t+1}}\beta_{t+1i}\left(\mathrm{P}_{t+1i}+\mathrm{b}_{t+1i}^{2}\right)-\left(\sum_{i=1}^{I_{t+1}}\beta_{t+1i}\mathrm{b}_{t+1i}\right)^{2},
\]
\[
\beta_{t+1i}=\frac{\phi\left(P_{A,t+1}-\mathrm{c}_{t+1i}-\gamma*P_{B,t+1},\mathrm{V}_{t+1t}\right)}{\sum_{i=1}^{I_{t+1}}\phi\left(P_{A,t+1}-\mathrm{c}_{t+1i}-\gamma*P_{B,t+1},\mathrm{V}_{t+1t}\right)}.
\]
\item Step 7: Repeat Step 2-6 for $t=0,1,2,...,T$.
\end{itemize}
$\small\{\bar{x}_{t}\small\}_{t=1}^{T}$ from Step 6 is our estimation of the spread. To estimate the unknown parameter in the model, we first write the
log-likelihood function as 
\begin{eqnarray*}
L_{T}^{G}\left(\psi\right) & \equiv & \sum_{t=0}^{T}\log f^{G}\left(\psi;P_{A,t},P_{B,t}\right)=\\
 & = & \sum_{t=1}^{T}\log\left[\sum_{i}^{I_{t+1}}\frac{1}{\sqrt{2\pi\mid\mathrm{V}_{t+1i}\mid}}\exp\left(-\frac{\left(P_{A,t+1}-\bar{P}_{A,t+1i}\right)^{2}}{2*\mathrm{V}_{t+1i}}\right)\right]
\end{eqnarray*}
and MLE of the unknow parameter would be determined to maximize the above likelihood, that
is,
\[
\hat{\psi}_{MLE}=\operatorname*{argmax}_{\psi\in\varPhi}L_{T}^{G}\left(\psi\right).
\]

\subsection{Benchmark Trading Strategies}

\begin{figure}
\centering
\caption{Trading Strategy A}
\includegraphics[width=16cm,height=10cm]{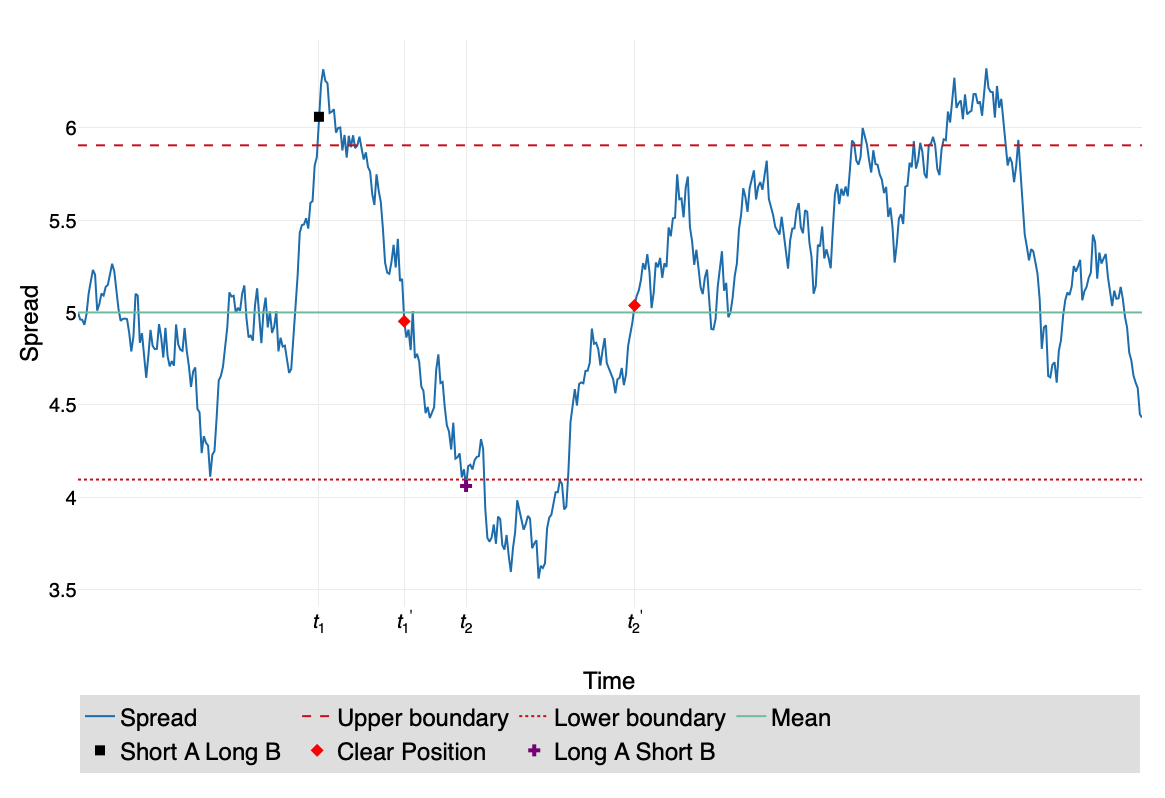}

\label{fig:strA}
\end{figure}

\begin{figure}
\centering
\caption{Trading Strategy B}
\includegraphics[width=16cm,height=10cm]{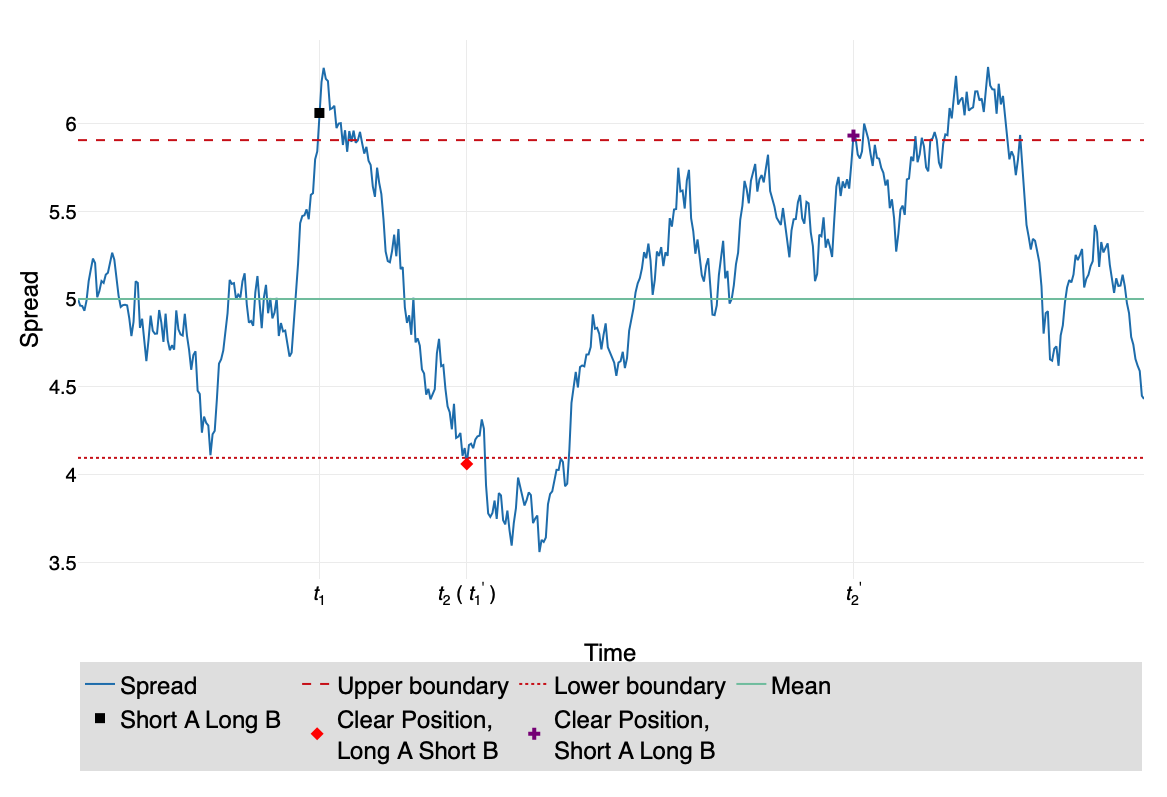}

\label{fig:strB}
\end{figure}

As we discussed in Section 1, the basic idea for pairs trading is to open a trade (short one asset and long the other one) when the spread deviates from the equilibrium and close the trading when the spread settle back to the equilibrium. The trading strategies for pairs trading are constructed based on this idea. We use Figure \ref{fig:strA} and Figure \ref{fig:strB} to illustrate two benchmark trading strategies
(hereafter Strategy A and Strategy B). In Figure \ref{fig:strA} and Figure \ref{fig:strB}, the same estimated spread is
plotted as solid lines, and a preset upper-boundary
$U$ and a preset lower-boundary $L$ are plotted as dashed lines.
We will discuss how to choose the optimal $U$ and $L$ in Section
3.2. The upper-boundary and lower-boundary act as thresholds
to determine whether the spread deviates from the long-term equilibrium enough, and we use these two criteria to open a trade. Also, a preset value $C$ acts
as a threshold to determine whether the spread settles back to the long-term equilibrium, and we use this criterion to close a trade. In this paper, we take $C$ as the mean of the
spread, and plot it as solid green line in both Figure \ref{fig:strA} and Figure \ref{fig:strB}. 

In Strategy A (illustrated in Figure \ref{fig:strA}), a trade is opened
at $t_{1}$ when the spread is higher than or equal to $U$. In this
case, we sell 1 share of stock A and buy $\gamma$ share of stock
B. At $t_{1}^{\prime}$ when the spread is less than or equal to the
mean (i.e., $C$), we close the trade and clear the position. The
return from this trade is thus $U-C$. At $t_{2}$ when the spread
is less than or equal to $L$, , we open a trade by buying 1 share
of stock A and sell $\gamma$ share of stock B. We close this trade
and clear the position at $t_{2}^{\prime}$ when the spread is higher
than or equal to the mean. The return from this trade is $C-L$.

In Strategy B (illustrated in Figure \ref{fig:strB}), we open a trade when the spread cross the upper-boundary
from below (e.g., at $t_{1}$ ) or cross the lower-boundary from above
(e.g., at $t_{2}$ ). Unlike the Strategy A, We will hold the portfolio
until we need to switch the position. Thus in Strategy B, we clear
the exposure at the same time when we open a new trade ( i.e., $t_{2}$
and $t_{1}^{\prime}$ coincide).

\subsection{Conventional Trading Rules}

In the implementation of pairs trading, trading rule for a specific trading strategy is the computation of optimal thresholds $U$and $L$ based on that strategy to fulfill an investing objective\footnote{Investing objective could be various, such as maximizing the expected cumulative return or maximizing the Sharpe ratio.}. There are three popular approaches for computing the optimal thresholds
$U$and $L$ when the model (\ref{eq:pt2}) is linear, homoscedastic and Gaussian (i.e., $f$ is linear, $g$ is a constant and $\eta$ is a Gaussian noise). The optimal trading rule for a general specification of model (\ref{eq:pt2}) will be given in Section 3.4.
\begin{itemize}
\item Rule $\textrm{I}$: Ad hoc boundaries
\end{itemize}
Rule $\textrm{I}$ takes $U$ to be one (1-$\sigma$
rule) or two (2-$\sigma$ rule) standard deviations above the mean, $L$ to be one or two standard deviations below the mean and
$C$ to be the mean of the spread. This rule is simple
and popular in practice. In particular, the 2-$\sigma$ rule was first applied by Gatev
et al. (2006) and later checked by Moura et al. (2016), Zeng and Lee
(2014) and Cummins and Bucca (2012). The 1-$\sigma$ rule was discussed
in Zeng and Lee (2014) and the performance of 1-$\sigma$ rule and
2-$\sigma$ rule was compared in the same paper.
\begin{itemize}
\item Rule II : Boundaries based on the first-passage-time
\end{itemize}
This rule was first adopted by Elliott et al. (2005) and later by
Moura et al. (2016). Suppose $Z_{t}$ follows
a standardized Ornstein--Uhlenbeck process: 
\[
dZ_{t}=-Z_{t}dt+\sqrt{2}dW_{t}
\]
Let $T_{0,Z_{0}}$ be the first passage time of $Z_{t}$:
\[
T_{0,Z_{0}}=\inf\{t\geq0,Z(t)=0|Z(0)=Z_{0}\}.
\]
$T_{0,Z_{0}}$ has a pdf known explicitly:
\[
f_{0,Z_{0}}(t)=\sqrt{\frac{2}{\pi}}\frac{|Z_{0}|e^{-t}}{\left(1-e^{-2t}\right)^{3/2}}\exp\left(-\frac{Z_{0}^{2}e^{-2t}}{2\left(1-e^{-2t}\right)}\right)
\]
$f_{0,Z_{0}}(t)$ can be maximized at $t^{*}$ given by:
\[
t^{*}=\frac{1}{2}\ln\left[1+\frac{1}{2}\left(\sqrt{\left(Z_{0}^{2}-3\right)^{2}+4Z_{0}^{2}}+Z_{0}^{2}-3\right)\right]
\]
Here $t^{*}$ is the most possible time, given the value of current spread, that the spread will settle back to the mean. In model (\ref{eq:pt2}), if the spread $x$ follows (discrete time) Ornstein--Uhlenbeck
process, then we can first standardize $x$, and then above formula
for $t^{*}$ can be used to construct the optimal $C$. Similar idea can be applied to compute the optimal upper-boundary $U$ and lower-boundary $L$. 
%The problem of this trading rule is that it is not optimizing the popular investing objectives such as the return or the Sharpe ratio.
\begin{itemize}
\item Rule III: Boundaries based on the renewal theorem
\end{itemize}
This rule was first proposed by Bertram (2010), and then extended by Zeng and
Lee (2014). In this rule, each trading cycle is separated into two parts, where $\tau_{1}$ can be used to denote the time
from taking (long or short) position to clearing the position, and
$\tau_{2}$ can be used to denote the time from clearing position to opening
next trading. That is,
\[
\tau_{1}=\inf\left\{ t;\hat{x}_{t}=C|\hat{x}_{0}=U\right\} 
\]
\[
\tau_{2}=\inf\left\{ t;\hat{x}_{t}=U|\hat{x}_{0}=C\right\} 
\]
Suppose $T$ is the total trading duration we have for a pair, and
$N_{T}$ is the number of transactions we can have in the period
$\left[0,T\right]$. Then, by the renewal theorem, the return
per unit time is given by:
\[
\left(U-C\right)\lim_{T\rightarrow\infty}\frac{E\left(N_{T}\right)}{T}=\frac{U-C}{E\left(\tau_{1}+\tau_{2}\right)}.
\]
where $E\left(\tau_{1}\right)$ and $E\left(\tau_{2}\right)$ can be computed
based on the density of first passage time,  mentioned in Rule II.

The problem of this rule is, as Zeng and Lee (2014) have pointed
out, that when there is no transaction cost, this strategy implies $U$
(and $L$) will be arbitrarily close to $C$. This implies that the trader
values the trading frequency more than the profit per trade. Consequently, this
could increase the risk of the portfolio significantly.

\subsection{The New Trading Strategy}

\begin{figure}[p]
\caption{Trading Strategy C}
\begin{subfigure}{1\textwidth}
  \centering
  \captionsetup{belowskip=12pt,aboveskip=4pt}
  \caption{Trading Strategy C in Homoscedastic Model}
  \includegraphics[width=16cm,height=10cm]{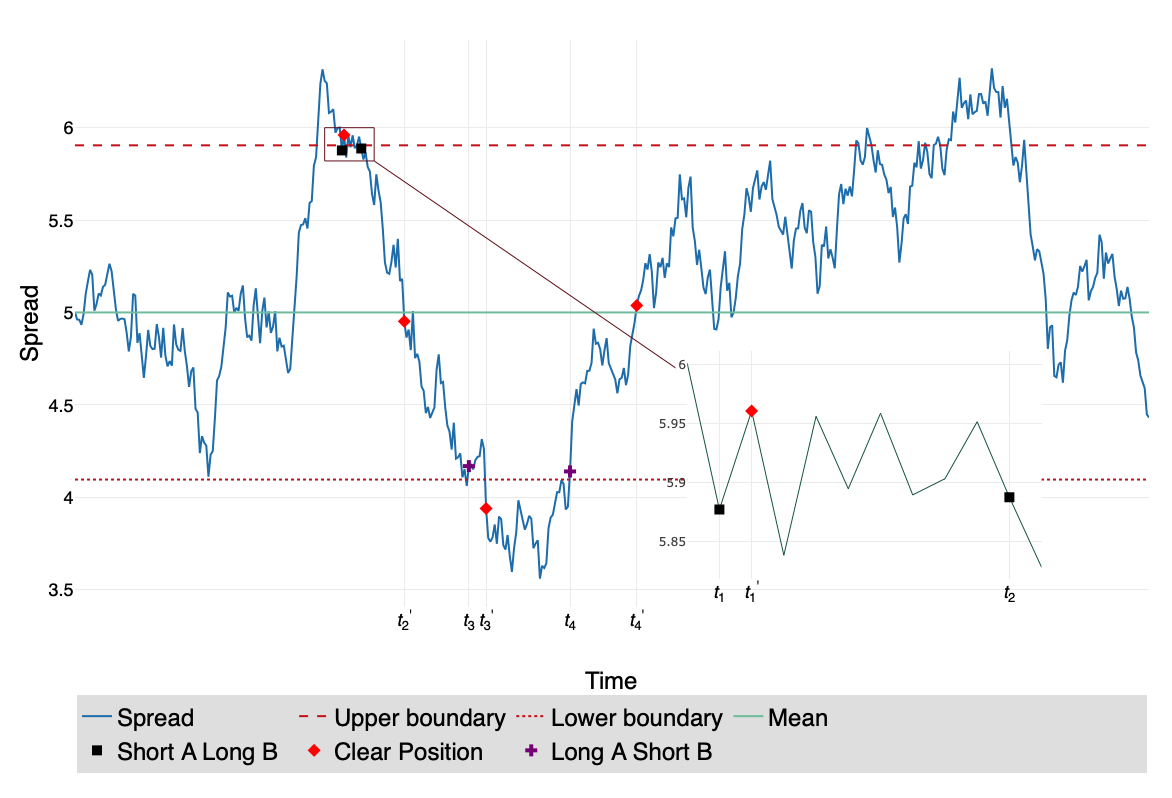}
  
  \label{fig:strC_1}
\end{subfigure}
\begin{subfigure}{1\textwidth}
  \centering
  \captionsetup{belowskip=12pt,aboveskip=4pt}
  \caption{Trading Strategy C in Heteroscedastic Model}
  \includegraphics[width=16cm,height=10cm]{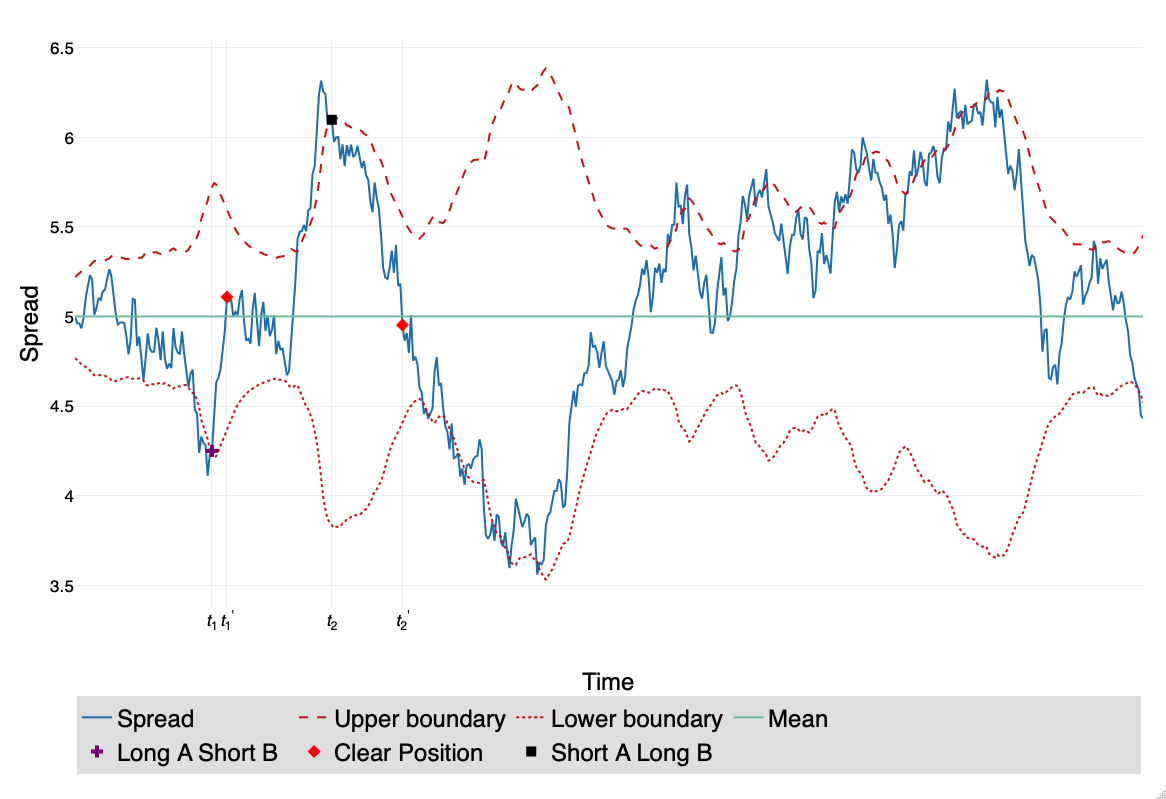}
  
  \label{fig:strC_2}
\end{subfigure}

\label{fig:strC}
\end{figure}

We summarize the new trading strategy (hereafter Strategy C) in Figure
\ref{fig:strC}. The basic idea of Strategy C is similar to both Strategy A and Strategy B: open a trade when the spread is far away from the equilibrium and close the trade when the spread settle back to the equilibrium. Unlike the Strategy A and B, in Strategy C, we open a trade when
the spread cross the upper-boundary from above (or cross the lower-boundary from below), and we clear the position when the spread cross
the mean, or cross the boundaries ($U$ and $L$) after a trade has been opened (i.e., the spread cross the upper-boundary from below or the lower-boundary from above). For example, in Figure \ref{fig:strC_1} for a homoscedastic model, at $t_{1}$, $t_{2}$, $t_{3}$ and $t_{4}$ we open a trade; and at $t_{1}^{\prime}$, $t_{2}^{\prime}$, $t_{3}^{\prime}$, and $t_{4}^{\prime}$ we clear the exposure. In Figure \ref{fig:strC_2} for a heteroscedastic model, we open a trade at $t_{1}$ and $t_{2}$; and we close the trade at $t_{1}^{\prime}$, and $t_{2}^{\prime}$.

We now discuss the properties of this trading strategy when the model (\ref{eq:pt2}) is homoscedastic (i.e., the $g$ function is constant) and when it is heteroscedastic (i.e., $g$ is a general function). In the first situation, the main benefit of Strategy C is that we can avoid holding the portfolio when the
spread is larger than the upper boundary (or smaller than the lower
boundary). This would significantly decrease the risk and drawdown
of the portfolio. The main drawback of Strategy C is that the return can be lower
because we open the trade when the spread is closer to the mean of the spread than in Strategy A. Therefore, there is a tradeoff between the risk and the return. In the situation when the model (\ref{eq:pt2}) is heteroscedastic, this strategy can not only reduce the risk, it can also improve the return.  This is because the opening of a trade now depends on the level of the volatility and, as a result, the boundaries are no longer constant over time. The logic of this new strategy is illustrated in Figure \ref{fig:strC_1} and \ref{fig:strC_2}, for homoscedastic and heteroscedastic cases, respectively.

%%%%%%%%%%%%%%%%%%%%%%%%%%%%%%%%%%%%%%%%%%%%%%

\subsection{Simulation Based Method for Optimal Trading Rule}

For a general specification of model (\ref{eq:pt1})-(\ref{eq:pt2}), the conventional trading rules in Section 3.2 are difficult to be applied. For example, the 1-$\sigma$ rule or 2-$\sigma$ rule cannot be applied when the model (\ref{eq:pt2}) is heteroscedastic; for a complicated specification of model (\ref{eq:pt2}), it's impossible to derive the density of the first passage time explicitly, thus Rule II and Rule III are unavailable in this case. 

To compute the optimal trading rule under model (\ref{eq:pt2}) for all of the trading strategies, we propose to select the optimal
boundaries ($U$ and $L$, we set $C$ as the mean of spread by default)
based on the Monte Carlo simulation of the spread (equation (\ref{eq:pt2})
given the estimation of the unknown parameters). Different criterion or investing objectives,
such as expected return, Sharpe ratio or Calmar ratio\footnote{Let $CR_{a,t}$ be the cumulative return of portfolio $a$ at time
$t$, and we define the maximum drawdown of the cumulative return
across time $0$ to $T$ as $MD_{a,T}$:
\[
MD_{a,T}=\sup_{t\in\left[0,T\right]}\left[\sup_{\tau\in\left[0,t\right]}CR_{a,\tau}-CR_{a,t}\right].
\]
Then the Calmar ratio can be defined in a similar way as the Sharpe
ratio:
\[
Calmar_{a}\equiv\frac{E\left(R_{a}\right)}{MD_{a,T}}
\]
where $E\left(R_{a}\right)$ is the expected return of portfolio $a$.} could be used to determine the optimal boundaries for a given trading strategy. 

Now we use the following four specifications of model (\ref{eq:pt2}) to describe the detail about the  computation of the new trading rules.

\begin{itemize}
\item Model 1: $x_{t+1}=0.9590*x_{t}+0.0049*\eta_{t}$, $\eta_{t}\sim N\left(0,1\right)$
\item Model 2: $x_{t+1}=0.9*x_{t}+0.5590*x_{t}^{2}+0.0049*\eta_{t}$, $\eta_{t}\sim N\left(0,1\right)$
\item Model 3: $x_{t+1}=0.9590*x_{t}+\sqrt{\left(0.00089+0.08*x_{t}^{2}\right)}*\eta_{t}$,
$\eta_{t}\sim N\left(0,1\right)$
\item Model 4: $x_{t+1}=0.9590*x_{t}+\frac{0.0049}{\sqrt{3}}*\eta_{t}$,
$\eta_{t}\sim t_{3}$
\end{itemize}

Model 1 is a linear, homoscedastic, and Gaussian model. This is the most popular model used for pairs trading. See Elliott et al. (2005) and Moura et al. (2016) for examples of this model. Model 2 is a nonlinear model, Model 3 is a heteroscedastic model, and Model 4 is a non-Gaussian model. The last three models are different extensions of Model 1 and have never been discussed in the literature on pairs trading. These four models can be considered as the benchmark models for pairs trading. Further extensions are available based on the combination of these four models, and our simulation based method for optimal trading rule can also be applied to them.

%More specifically, for Model 1 and Model 4, the boundaries are selected
%as the ratio to $0.0173$ \footnote{This is $0.0049/\sqrt{1-0.9590^{2}}$.};
%for Model 2, they are selected as the ratio to 0.011 \footnote{This is based on sample standard deviation of simulation of Model 2.};
%for Model 3, they are selected as the ratio to $\left(\sqrt{0.00089+0.08*\left(\bar{x}_{t}\right)^{2}}\right)/\sqrt{1-0.9590^{2}}$.

For every specification of Model 1-4, we will calculate the optimal trading rules through the $N$ simulations of the spread for Strategy A, B and C respectively, and compare the resulting performances of the three strategies based on the expected return, Sharpe ratio. More specifically, across all of the examples, we represent the optimal trading rule (upper-boundary $U$ and lower-boundary $L$) as the ratio to one standard deviation of the spread, and we consider the upper-boundary $U$ between $\left[0.1,2.5\right]$ and lower-boundary $L$
between $\left[-2.5,-0.1\right]$ for a grid size of 0.1. For every specification of Model 1-4 and every realization of the process of the spread $\small\{ x_{t}^{\left(m,n\right)}\small\} _{t=0}^{T}$, where $m=1,2,3,4; n=1,\ldots N$, we choose $U_{i}$ from $\left[0.1,2.5\right]$ and $L_{j}$ from
$\left[-2.5,-0.1\right]$, where $i,j=1,...,25$, and compute the resulting cumulative return and Sharpe
ratio for difference strategies. More specifically, We denote the cumulative return
and Sharpe ratio as $CR_{i,j}^{m,k,n}$ and $SR_{i,j}^{m,k,n}$ respectively,
where $m$ is for different models, $k$ is for difference strategies
and $n$ is for different realization of the spread in simulation. For Model $m$
and strategy $k$, the resulting expected cumulative return $CR_{i,j}^{m,k}$
and Sharpe ratio $SR_{i,j}^{m,k}$ are computed as 
\[
CR_{i,j}^{m,k}=\frac{1}{N}\sum_{n=1}^{N}CR_{i,j}^{m,k,n}
\]
\[
SR_{i,j}^{m,k}=\frac{1}{N}\sum_{n=1}^{N}SR_{i,j}^{m,k,n}.
\]
Then the optimal trading rule ($U_{m,k}^{*}$, $L_{m,k}^{*}$) is
selected to maximize $CR_{i,j}^{m,k}$ or $SR_{i,j}^{m,k}$, that
is,
\begin{eqnarray*}
\left[U_{m,k}^{*},L_{m,k}^{*}\right] & = & \arg\max_{U_{i},L_{j}}z_{i,j}^{m,k}
\end{eqnarray*}
where $z=CR$ or $SR$. Across all of the examples, we set the total trading period to be 1000 trading days (or approximately four years), and we set the simulation size to be $N=10000$. For
simplicity, we assume the transaction cost is 20 bp (0.2\%) \footnote{This transaction cost is on one asset of the pair. Since a complete trading includes transactions on two assets, the total transaction cost of one complete trading is 40 bp.}, and annualized risk free rate is set to be 0.

In Table 1, we report the optimal trading
rule for every combination of the 4 models and 3 strategies, and the
resulting expected cumulative return and Sharpe ratio\footnote{If the spread and the strategy is symmetric around the mean, then
the optimal upper boundary and lower boundary should also be symmetric
around the mean, i.e, $U^{*}=-L^{*}$. However, due to the approximation
error in gridding, the absolute values of $U^{*}$ and $L^{*}$ may
not be exactly the same in Table 1.}. As we can find from this table, Strategy C outperforms other two strategies when the model is heteroscedastic in both the cumulative return and the Sharpe ratio; also, for other homoscedastic models (Model 1, 2 and 4), the Sharpe ratio of Strategy C is competitive, although the cumulative return is not. This supports our discussion of this new strategy in Section 3.3.

We leave the detailed results of simulation method in appendix. More precisely,
the expected cumulative returns and Sharpe ratio as functions of various choices of $U$ and $L$ are given in Figure \ref{fig:model1}-\ref{fig:model4} for every possible combination of the three strategies and four models. The return is displayed in number, not in percentage through all figures.

\begin{table*}[ht]
\caption{Optimal selection of trading rule for cumulative return and Sharpe ratio}
\centering
\begin{threeparttable}

\begin{tabular}{c|c|ccc|ccc}
\hline
Model & Strategy & $U^{*}$ & $L^{*}$ & CR & $U^{*}$ & $L^{*}$ & SR\\
\hline\hline
\multirow{3}{5em}{Model 1} & A & 0.7 & -0.7 & 0.2508 & 1.1 & -1 & 0.0573\\
 & B & 0.5 & -0.5 & 0.2745 & 0.5 & -0.5 & 0.0522\\
 & C & 1 & -1 & 0.1934 & 0.9 & -0.9 & 0.0679\\
\hline 
\multirow{3}{5em}{Model 2} & A & 0.8 & -0.8 & 0.2749 & 1.2 & -1.3 & 0.1302\\
 & B & 0.6 & -0.6 & 0.3016 & 0.6 & -0.6 & 0.1198\\
 & C & 1.2 & -1.3 & 0.1640 & 1.2 & -1.3 & 0.1162\\
\hline 
\multirow{3}{5em}{Model 3} & A & 0.3 & -0.2 & 3.9413 & 0.4 & -0.4 & 0.0751\\
 & B & 0.1 & -0.1 & 4.0139 & 0.1 & -0.1 & 0.0743\\
 & C & 0.8 & -0.8 & 6.6763 & 0.1 & -0.1 & 0.2499\\
\hline 
\multirow{3}{5em}{Model 4} & A & 0.6 & -0.6 & 0.3792 & 1 & -1 & 0.0881\\
 & B & 0.4 & -0.5 & 0.4071 & 0.5 & 0.5 & 0.0782\\
 & C & 1 & -1 & 0.2243 & 1 & -1 & 0.0829\\
 \hline
\end{tabular}
\begin{tablenotes}
\footnotesize
\item Note: The third and forth columns are the optimal upper-boundary and lower-boundary based on maximizing the cumulative return, and the fifth column is the resulting cumulative return. The sixth and seventh columns are the optimal upper-boundary and lower-boundary based on maximizing the Sharpe ratio, and the eighth column is the resulting Sharpe ratio.  The cumulative return is displayed in number, not in percentage.
\end{tablenotes}
\end{threeparttable}
\label{table_sim}
\end{table*}

\subsection{Summary}

We are now in a position to summarize the procedure for pairs trading based on model (\ref{eq:pt1})-(\ref{eq:pt2})
and conclude this section.
\begin{itemize}
\item Step 1: Choose a specific model for (\ref{eq:pt1})-(\ref{eq:pt2}). Given
this model and observations $\small\{ P_{A,t},P_{B,t}\small\} _{t=0}^{T}$,
we run Quasi Monte Carlo Kalman filter and get the filtered estimation
of the spread $\small\{ \bar{x}_{t}\small\} _{t=0}^{T}$ and the estimation
of the unknown parameter $\hat{\psi}$ in the model. The
detail of running QMCKF has been discussed in Section 3.1.
\item Step 2: Choose a trading strategy, and determine the optimal trading
rule (the optimal $U$ and $L$) for a specific criterion based on
Monte Carlo simulation based on the data until time $T$. The detail of this step can be found in Section 3.2-3.5.
\item Step 3: For $t>T$, we run QMCKF and estimate $\bar{x}_{t}$ with ${\psi}=\hat{\psi}$, the estimate of the parameter we get in Step 1 . We use this $\small\{ \bar{x}_{t}\small\} _{t>T}$
and follow the preset trading strategy and optimal trading rule from
Step 2 to generate the trading signal for trading.
\end{itemize}

\section{Applications}

In this section, we test the performance of Pairs Trading through
nonlinear and non-Gaussian state space modeling for different trading
strategies. Across all of the applications in this section, we assume
the transaction cost is 20 bp and the annualized risk free rate is
2\%, and we test the performance of Strategy A, B and C for two specifications
of model (\ref{eq:pt2}):
\begin{itemize}
\item Model $\textrm{I}$: $x_{t+1}=\theta_{0}+\theta_{1}x_{t}+\theta_{2}*\eta_{t}$,
$\eta_{t}\sim{N}\left(0,1\right)$
\item Model $\textrm{II}$: $x_{t+1}=\theta_{0}+\theta_{1}x_{t}+\sqrt{\theta_{2}+\theta_{3}x_{t}^{2}}*\eta_{t}$,
$\eta_{t}\sim{N}\left(0,1\right)$
\end{itemize}

\subsection{Pepsi vs Coca}

In this example, we examine the performance of Pairs Trading for PEP
(Pepsi) and KO (Coca). The data is the daily observation of adjusted
closing prices of PEP and KO from 01/03/2012-06/28/2019.

Table \ref{para_pepko} reports the parameter estimation of both Model $\textrm{I}$ and Model
$\textrm{II}$ for this pair. The trading signal for Model $\textrm{I}$ is given in Figure
\ref{pepko_signal_m1} and that for Model $\textrm{II}$ is given in Figure \ref{pepko_signal_m2}, and the annualized performance
(annualized return, annualized Std Dev, annualized Sharpe ratio and
Calmar ratio, and annualized Pain index) is given in Table \ref{pepko}. The
plot of the cumulative return and drawdown of every strategy through the whole trading period for both models are given in Figure \ref{pepko_perf_m1} and \ref{pepko_perf_m2}. It's easy to find that in Model $\textrm{II}$, the annualized
return of Strategy C is almost 50\% higher than those of Strategy
A and B, while Strategy C keeps the risk (measured by Annualized Std
Dev) almost half of Strategy A or B. By comparing the Sharpe ratio,
Calmar ratio and Pain index, we can find this improvement is significant.
While the difference of performances of Strategy A and Strategy B
across the two models is limited. This implies the effect of heteroskedasticity
modelling to the performances of Strategy A and B is not significant.
This is because in Strategy A and B, the hedging portfolio will be
held until the spread is around the mean, so the frequency of changing
positions is low in Strategy A or B than that in Strategy C. This
can be easily confirmed by counting the trading numbers based on Figure
\ref{pepko_signal_m1} and Figure \ref{pepko_signal_m2}.

\begin{table}[h!]
\centering
\caption{Parameter estimation of Model $\textrm{I}$ and Model $\textrm{II}$ on PEP vs KO}
\begin{threeparttable}
\begin{tabular}{c|c|c}
\hline
 & Model $\textrm{I}$ & Model $\textrm{II}$\\
\hline \hline
$\gamma$ & 1.98 & 2.03\\
$\sigma_{\varepsilon}^{2}$ & 0.012 & 0.011\\
$\theta_{0}$ & -0.0001 & -0.001\\
$\theta_{1}$ & 0.9572 & 0.9330\\
$\theta_{2}$ & 0.029 & 0.0003\\
$\theta_{3}$ & - & 0.1283\\
\hline
\end{tabular}
\end{threeparttable}

\label{para_pepko}
\end{table}

\begin{table}[h!]
\centering
\caption{Annualized Performance of Pairs Trading on PEP vs KO}
\begin{threeparttable}
\begin{tabular}{c|c|c|c|c|c}
\hline
  & Return & Std Dev & Sharpe  & Calmar  & Pain index\tabularnewline
\hline \hline
Strategy A, Model $\textrm{I}$ & 0.1311 & 0.0988 & 1.1003 & 1.3742 & 0.0195\tabularnewline
Strategy B, Model $\textrm{I}$ & 0.1385 & 0.1153 & 1.0052 & 1.2204 & 0.0334\tabularnewline
Strategy C, Model $\textrm{I}$ & 0.0618 & 0.0534 & 0.7649 & 0.8243 & 0.0087\tabularnewline
Strategy A, Model $\textrm{II}$ & 0.1340 & 0.1038 & 1.0751 & 1.4040 & 0.0200\tabularnewline
Strategy B, Model $\textrm{II}$ & 0.1407 & 0.1139 & 1.0366 & 1.2398 & 0.0258\tabularnewline
Strategy C, Model $\textrm{II}$ & 0.2186 & 0.0659 & 2.9518 & 8.2384 & 0.0030\tabularnewline
\hline
\end{tabular}
\begin{tablenotes}
\footnotesize
\item Note: The data is from 01/03/2012-06/28/2019. The return is displayed in number, instead of in percentage.
\end{tablenotes}
\end{threeparttable}

\label{pepko}
\end{table}

\subsection{EWT vs EWH}

In this example, we examine the performance of Pairs Trading for EWT
and EWH. The data is the daily observation of adjusted closing prices
of EWT and EWH from 01/01/2012-05/01/2019. EWT is the iShares MSCI
Taiwan ETF managed by BlackRock, which seeks to track the investment
results of an index composed of Taiwanese equities, and EWH is that
for Hong Kong equities. Following the example of PEP vs KO, we will
test the performance of Strategy A, B and C for Model $\textrm{I}$ and Model
$\textrm{II}$. We report the parameter estimation in Table \ref{para_ewtewh} and the trading signal
in Figure \ref{ewtewh_signal_m1} and Figure \ref{ewtewh_signal_m2}. By comparing the annualized performance
in Table \ref{ewtewh}, we can find the heteroskedasticity modeling can improve
the performance of Strategy C significantly, while has no effect on
Strategy A or B. Also, the riskiness of Strategy B (small Sharpe ratio
and Calmar ratio and high annualized standard variance) is confirmed
again in this example. We also 
plot the cumulative return and drawdown of every strategy through the whole trading period for both models in Figure \ref{ewtewh_perf_m1} and \ref{ewtewh_perf_m2}.

\begin{table}[h!]
\centering
\caption{Parameter estimation of Model $\textrm{I}$ and Model $\textrm{II}$ on EWT vs EWH}
\begin{threeparttable}
\begin{tabular}{c|c|c}
\hline
 & Model $\textrm{I}$ & Model $\textrm{II}$\tabularnewline
\hline \hline
$\gamma$ & 1.40 & 1.42\tabularnewline
$\sigma_{\varepsilon}^{2}$ & 0.0007 & 0.0006\tabularnewline
$\theta_{0}$ & -0.0004 & -0.0015\tabularnewline
$\theta_{1}$ & 0.9898 & 0.9589\tabularnewline
$\theta_{2}$ & 0.0337 & 0.0016\tabularnewline
$\theta_{3}$ & - & 0.1136\tabularnewline
\hline
\end{tabular}
\end{threeparttable}

\label{para_ewtewh}
\end{table}

\begin{table}[h!]
\centering
\caption{Annualized Performance of Pairs Trading on EWT vs EWH}
\begin{threeparttable}
\begin{tabular}{c|c|c|c|c|c}
\hline
  & Return & Std Dev & Sharpe  & Calmar  & Pain index\tabularnewline
\hline \hline
Strategy A, Model $\textrm{I}$ & 0.1480 & 0.1111 & 1.1277 & 1.3042 & 0.0156\tabularnewline
Strategy B, Model $\textrm{I}$ & 0.1109 & 0.1362 & 0.6531 & 0.7836 & 0.0328\tabularnewline
Strategy C, Model $\textrm{I}$ & 0.1294 & 0.0740 & 1.4458 & 3.0926 & 0.0080\tabularnewline
Strategy A, Model $\textrm{II}$ & 0.1402 & 0.1223 & 0.9622 & 1.2354 & 0.0196\tabularnewline
Strategy B, Model $\textrm{II}$ & 0.1093 & 0.1349 & 0.6473 & 0.7717 & 0.0306\tabularnewline
Strategy C, Model $\textrm{II}$ & 0.3184 & 0.0752 & 3.8892 & 10.3005 & 0.0032\tabularnewline
\hline
\end{tabular}
\begin{tablenotes}
\footnotesize
\item Note: The data is from 01/03/2012-06/28/2019. The return is displayed in number, instead of in percentage.
\end{tablenotes}
\end{threeparttable}

\label{ewtewh}
\end{table}

\subsection{Pairs Trading on US Banks Listed on NYSE}

We use this example to illustrate the improvement of our new modelling and strategy by implementing pairs trading on US banks listed on NYSE during 01/01/2013-01/10/2019. To avoid data snooping and make our results more concrete, we use a simple way to choose assets and construct pairs. More precisely, based on the market capacity, we select the 5 largest banks to construct
the group of large banks and the 5 smallest banks to construct the group
of small banks. The large bank group includes: JPM, BAC, WFC, C and
USB\footnote{JPM is for J P Morgan Chase \& Co; BAC is for Bank of America Corporation; WFC is for Wells Fargo \& Company; C is for Citigroup Inc.; USB is for U.S. Bancorp.}
, and the small bank group includes: CPF, BANC, CUBI, NBHC, FCF\footnote{CPF is for CPB Inc.; BANC is for Banc of California, Inc.; CUBI is for Customers Bancorp, Inc.; NBHC is for National Bank Holdings Corporation; FCF is for First Commonwealth Financial Corporation.}.
We compare the performance between Model
$\textrm{I}$ combined with Strategy A and Model $\textrm{II}$ combined with Strategy C. Model
$\textrm{I}$ combined with Strategy A is a popular approach in the existing literature on pairs trading, and it can be a good benchmark for comparison.

In Table \ref{bigsmall_big}, we report the performance of these two approaches on 10
pairs among the large banks. The performance on 10 pairs among the small banks is given in Table \ref{bigsmall_small}. It's easy to find that Model $\textrm{II}$ combined with Strategy C outperforms
Model
$\textrm{I}$ combined with Strategy A through almost all of the pairs, either in the sense of annualized
return or annualized Sharpe ratio. And the improvement of Model $\textrm{II}$ combined with Strategy C in Sharpe ratio is much more significant than that in return. For example, when trading is implemented on pairs among large banks, the improvement on return is 41.29\%, and the improvement on Sharpe ratio is 89.23\%; and if trading is implemented on pairs among small banks, the improvement on return is 74.41\%, and the improvement on Sharpe ratio is 151.8\%.

Also, by comparing the results in Table \ref{bigsmall_big} and \ref{bigsmall_small}, we can find that the performance of pairs among small banks would be better than that among large banks, either Model
$\textrm{I}$ combined with Strategy A or Model $\textrm{II}$ combined with Strategy C is applied for trading. For example, if we exercise Model
$\textrm{I}$ combined with Strategy A, the mean of returns of all pairs among large banks would be 0.0703, that among small banks can be improved to 0.1524; and if Model $\textrm{II}$ combined with Strategy C is exercised, we could get an improvement of 0.1664 (from 0.0994 to 0.2658) by switching from trading on large banks to trading on small banks. This is because the movement of prices of small banks is more volatile than that of large banks, and thus the volatility of the spread between small banks is bigger than that between large banks.

In Table \ref{bigsmall_all}, we report the performance of the two approaches of pairs trading on all possible pairs between large banks and small banks, that is, we pair one large bank with one small bank. For some pairs, such as JPM/CUBI and BAC/CUBI, the resulting spread is far from mean-reverting, thus the performance of pairs trading is poor for these pairs. Similiar to our findings from Table \ref{bigsmall_big} and \ref{bigsmall_small}, in this exercise, we can also find that the improvement of Model $\textrm{II}$ combined with Strategy C with respect to Model
$\textrm{I}$ combined with Strategy A on Sharpe ratio would be more significant than return (208.4\% on Sharpe ratio, and 103.6\% on return).

The results of Table \ref{bigsmall_big}-\ref{bigsmall_all} are also plotted in Figure \ref{fig:inter} and \ref{fig:intra} to give a more straightforward comparison of the performances. 

To further investigate the performance of pairs trading, we check the out-of-sample performance of the two approaches on the 10 bank stocks. More precisely, we separate 01/10/2012-01/12/2019 into two periods: 01/10/2012-01/01/2018 as in-sample period and 01/01/2018-01/12/2019 as out-of-sample period. We use the in-sample data to train the model, estimate the parameter of the model, and determine the optimal trading rules. In out-of-sample period, we use the parameters and optimal trading rules based on in-sample data to generate the trading signal. The results are given in Table \ref{bigsmallout_big_in}-\ref{bigsmallout_all_out}. We can confirm our earlier findings through these tables also: (1) Model $\textrm{II}$ combined with Strategy C outperforms Model
$\textrm{I}$ combined with Strategy A in both return and Sharpe ratio, and the improvement is more significant in Sharpe ratio. (2) The performance of pairs trading on small banks would be better than large banks. Also, by comparing the performance through in-sample period to out-of-sample period, we can find that pairing large bank with small bank would be more robust than pairing large banks only or small banks only. 

\section{Conclusion}

Pairs trading is a statistical arbitrage involves the long/short position
of overpriced and underpriced assets. Our result in this paper shows
that digging into the modeling and trading strategy can improve the
performance of pairs trading significantly and implies the great potential
of pairs trading on financial market. This can help the empirical
research on the general profitability of pairs trading and discussion
on the tests of market efficiency, and we leave this for future research.

\newpage
\appendix
\setcounter{table}{0}
\setcounter{figure}{0}
\renewcommand{\thetable}{A\arabic{table}}
\renewcommand{\thefigure}{A\arabic{figure}}

\clearpage

%\section{Tables}

\begin{sidewaystable}[h!]
\centering
\caption{Performance of Pairs Trading on Intergroup Pairs of Big Banks}

\begin{threeparttable}
\begin{tabular}{c|c|c|c|c|c|c|c|c}
\hline 
\multirow{2}{1.5em}{Pair} & \multirow{2}{4.5em}{Stock \#1} & \multirow{2}{4.5em}{Stock \#2} & \multicolumn{2}{c|}{Model $\textrm{I}$ + Strategy A} & \multicolumn{2}{c|}{Model $\textrm{II}$ + Strategy C} &  \multicolumn{2}{c}{Improvement (in \%)}\tabularnewline
\cline{4-9} \cline{5-9} \cline{6-9} \cline{7-9} \cline{8-9} \cline{9-9} 
 &  &  & {$\,\,\,\,$Return$\,\,\,\,$} & {Sharpe} & {$\,\,\,\,$Return$\,\,\,\,$} & {Sharpe}& {$\,\,\,\,$Return$\,\,\,\,$} & {$\,$Sharpe$\,$} \tabularnewline
\hline \hline 

1 & JPM & BAC & 0.1185 & 1.0030 & 0.0961 & 1.1126 & -18.90 & 10.93\tabularnewline
%\hline 
2 & JPM & WFC & 0.0229 & 0.2268 & 0.0581 & 0.7434 & 153.7 & 227.8\tabularnewline
%\hline 
3 & JPM & C & 0.0567 & 0.5359 & 0.1049 & 1.3486 & 85.01 & 151.7 \tabularnewline
%\hline 
4 & JPM & USB & 0.0412 & 0.3971 & 0.0663 & 0.7832 & 60.92 & 97.23 \tabularnewline
%\hline 
5 & BAC & WFC & 0.0451 & 0.3455 & 0.0695 & 0.6046 & 54.10 & 74.99 \tabularnewline
%\hline 
6 & BAC & C & 0.0874 & 0.8158 & 0.1369 & 1.7516 & 56.64 & 114.7\tabularnewline
%\hline 
7 & BAC & USB & 0.0554 & 0.3786 & 0.0923 & 1.0077 & 66.61 & 166.2 \tabularnewline
%\hline 
8 & WFC & C & 0.1031 & 0.8041 & 0.1014 & 0.9731 & -1.649 & 21.02\tabularnewline
%\hline 
9 & WFC & USB & 0.0591 & 0.5631 & 0.0674 & 0.8934 & 14.04 & 58.66 \tabularnewline
%\hline 
10 & C & USB & 0.1140 & 0.9040 & 0.2009 & 2.0862 & 76.23 & 130.8\tabularnewline
\hline 
\multicolumn{3}{c|}{Mean} & 0.0703 & 0.5974 & 0.0994 & 1.1304 & 41.29 & 89.23 \tabularnewline
%\hline 
\multicolumn{3}{c|}{Min} & 0.0229 & 0.2268 & 0.0581 & 0.6046 & 153.7 & 166.6\tabularnewline
%\hline 
\multicolumn{3}{c|}{Max} & 0.1185 & 1.0030 & 0.2009 & 2.0862 & 69.54 & 108.0\tabularnewline
%\hline 
\multicolumn{3}{c|}{Median} & 0.0579 & 0.5495 & 0.0942 & 0.9904 & 62.69 & 80.24\tabularnewline
\hline 
\end{tabular}
\begin{tablenotes}
\footnotesize
\item Note: Return is the annualized return, displayed in number, not in percentage. Sharpe is the annualized Sharpe ratio. Improvement is defined as $\frac{(\mathrm{Model\: \textrm{II}+Strategy\:C})-(\mathrm{Model\: \textrm{I}+Strategy\:A})}{|\mathrm{Model\:\textrm{I}+Strategy\:A}|}$
for return and Sharpe ratio respectively, measured in percentage.
\end{tablenotes}
\end{threeparttable}

\label{bigsmall_big}
\end{sidewaystable}

\begin{sidewaystable}[h!]
\centering
\caption{Performance of Pairs Trading on Intergroup Pairs of Small Banks}

\begin{threeparttable}
\begin{tabular}{c|c|c|c|c|c|c|c|c}
\hline 
\multirow{2}{1.5em}{Pair} & \multirow{2}{4.5em}{Stock \#1} & \multirow{2}{4.5em}{Stock \#2} & \multicolumn{2}{c|}{Model $\textrm{I}$ + Strategy A} & \multicolumn{2}{c|}{Model $\textrm{II}$ + Strategy C} &  \multicolumn{2}{c}{Improvement (in \%)}\tabularnewline
\cline{4-9} \cline{5-9} \cline{6-9} \cline{7-9} \cline{8-9} \cline{9-9} 
 &  &  & {$\,\,\,\,$Return$\,\,\,\,$} & {Sharpe} & {$\,\,\,\,$Return$\,\,\,\,$} & {Sharpe}& {$\,\,\,\,$Return$\,\,\,\,$} & {$\,$Sharpe$\,$} \tabularnewline
\hline \hline 

1 & CPF & BANC & 0.1832 & 0.6745 & 0.2158 & 1.3428 & 17.79 & 99.08 \tabularnewline
%\hline 
2 & CPF & CUBI & 0.1092 & 0.4736 & 0.2374 & 1.3563 & 117.4 & 186.4 \tabularnewline
%\hline 
3 & CPF & NBHC & 0.1436 & 0.7694 & 0.1912 & 1.2573 & 33.15 & 63.41  \tabularnewline
%\hline 
4 & CPF & FCF & 0.1162 & 0.7127 & 0.2175 & 1.7210 & 87.18 & 141.5\tabularnewline
%\hline 
5 & BANC & CUBI & 0.1583 & 0.5199 & 0.4820 & 1.9742 & 204.5 & 279.7\tabularnewline
%\hline 
6 & BANC & NBHC & 0.2105 & 0.8353 & 0.1807 & 1.1435 & -14.16 & 36.90\tabularnewline
%\hline 
7 & BANC & FCF & 0.1669 & 0.5830 & 0.3094 & 2.1898 & 85.38 & 275.6 \tabularnewline
%\hline 
8 & CUBI & NBHC & 0.1575 & 0.6049 & 0.2392 & 1.4485 & 51.87 & 139.5\tabularnewline
%\hline 
9 & CUBI & FCF & 0.1362 & 0.5593 & 0.2718 & 1.5292 & 99.56 & 173.4\tabularnewline
%\hline 
10 & NBHC & FCF & 0.1425 & 0.8161 & 0.3132 & 2.5273 & 119.8 & 209.7\tabularnewline
\hline 
\multicolumn{3}{c|}{Mean} & 0.1524 & 0.6549 & 0.2658 & 1.6490 & 74.41 & 151.8\tabularnewline
%\hline 
\multicolumn{3}{c|}{Min} & 0.1092 & 0.4736 & 0.1807 & 1.1435 & 65.48 & 141.4 \tabularnewline
%\hline 
\multicolumn{3}{c|}{Max} & 0.2105 & 0.8353 & 0.4820 & 2.5273 & 129.0 & 202.6 \tabularnewline
%\hline 
\multicolumn{3}{c|}{Median} & 0.1506 & 0.6397 & 0.2383 & 1.4889 & 58.29 & 132.7\tabularnewline
\hline 

\end{tabular}
\begin{tablenotes}
\footnotesize
\item Note: Return is the annualized return, displayed in number, not in percentage. Sharpe is the annualized Sharpe ratio. Improvement is defined as that in Table \ref{bigsmall_big}
\end{tablenotes}
\end{threeparttable}

\label{bigsmall_small}
\end{sidewaystable}

\begin{table}[h!]
\centering
\caption{Performance of Pairs Trading on Intragroup Pairs.}

\begin{threeparttable}
\begin{tabular}{c|c|c|c|c|c|c|c|c}
\hline 
\multirow{2}{1.5em}{Pair} & \multirow{2}{3.2em}{Stock \#1} & \multirow{2}{3.2em}{Stock \#2} & \multicolumn{2}{c|}{Model $\textrm{I}$ + Strategy A} & \multicolumn{2}{c|}{Model $\textrm{II}$ + Strategy C} & \multicolumn{2}{c}{Improvement (in \%)}\tabularnewline
\cline{4-9} \cline{5-9} \cline{6-9} \cline{7-9} \cline{8-9} \cline{9-9} 
 &  &  & {$\,\,\,\,$Return$\,\,\,\,$} & {Sharpe} & {$\,\,\,\,$Return$\,\,\,\,$} & {Sharpe}& {$\,\,\,\,$Return$\,\,\,\,$} & {$\,$Sharpe$\,$} \tabularnewline
 \hline\hline

1 & JPM & CPF & 0.0670 & 0.3965 & 0.1833 & 1.4799 & 173.6 & 273.2\\[-0.5ex]
%\hline 
2 & JPM & BANC & 0.0587 & 0.2396 & 0.0935 & 0.8334 & 59.28 & 247.8 \\[-0.5ex]
%\hline 
3 & JPM & CUBI & -0.0604 & -0.2669 & 0.0423 & 0.3536 & 170.0 & 232.5\\[-0.5ex]
%\hline 
4 & JPM & NBHC & 0.1860 & 0.9750 & 0.2683 & 2.1385 & 44.25 & 119.3\\[-0.5ex]
%\hline 
5 & JPM & FCF & 0.1151 & 0.7230 & 0.2594 & 2.3479 & 125.4 & 224.7\\[-0.5ex]
%\hline 
6 & BAC & CPF & 0.0778 & 0.3770 & 0.2486 & 1.5596 & 219.5 & 313.7 \\[-0.5ex]
%\hline 
7 & BAC & BANC & 0.0565 & 0.2124 & 0.1383 & 0.7916 & 144.8 & 272.7\\[-0.5ex]
%\hline 
8 & BAC & CUBI & -0.0959 & -0.3612 & 0.0473 & 0.5852 & 149.4 & 262.0 \\[-0.5ex]
%\hline 
9 & BAC & NBHC & 0.1942 & 0.9496 & 0.3420 & 2.4948 & 76.11 & 162.7 \\[-0.5ex]
%\hline 
10 & BAC & FCF & 0.1729 & 0.9061 & 0.2541 & 2.1954 & 46.96 & 142.3\\[-0.5ex]
%\hline 
11 & WFC & CPF & 0.0420 & 0.2149 & 0.1138 & 1.2746 & 171.0 & 493.1\\[-0.5ex]
%\hline 
12 & WFC & BANC & 0.1671 & 0.6058 & 0.2071 & 1.0214 & 23.94 & 68.60\\[-0.5ex]
%\hline 
13 & WFC & CUBI & 0.0606 & 0.2572 & 0.2053 & 1.3002 & 238.8 & 405.5\\[-0.5ex]
%\hline 
14 & WFC & NBHC & 0.1410 & 0.7844 & 0.1237 & 0.9464 & -12.27 & 20.65\\[-0.5ex]
%\hline 
15 & WFC & FCF & 0.1058 & 0.5948 & 0.1366 & 1.3104 & 29.11 & 120.3\\[-0.5ex]
%\hline 
16 & C & CPF & 0.1421 & 0.7000 & 0.2214 & 2.1513 & 55.81 & 207.3\\[-0.5ex]
%\hline 
17 & C & BANC & 0.0244 & 0.0961 & 0.1999 & 1.1101 & 719.3 & 1055\\[-0.5ex]
%\hline 
18 & C & CUBI & -0.0031 & -0.0138 & 0.0617 & 0.4357 & 2090 & 3257\\[-0.5ex]
%\hline 
19 & C & NBHC & 0.2164 & 1.0536 & 0.2927 & 2.3896 & 35.26 & 126.8\\[-0.5ex]
%\hline 
20 & C & FCF & 0.1520 & 0.7687 & 0.2246 & 1.8611 & 47.76 & 142.1\\[-0.5ex]
%\hline 
21 & USB & CPF & 0.0782 & 0.4494 & 0.2408 & 2.0902 & 207.9 & 365.1\\[-0.5ex]
%\hline 
22 & USB & BANC & 0.1435 & 0.5450 & 0.2361 & 1.7444 & 64.53 & 220.1\\[-0.5ex]
%\hline 
23 & USB & CUBI & -0.0678 & -0.2938 & 0.0700 & 0.3497 & 203.2 & 219.0\\[-0.5ex]
%\hline 
24 & USB & NBHC & 0.1911 & 1.2574 & 0.2384 & 2.1422 & 24.74 & 70.37\\[-0.5ex]
%\hline 
25 & USB & FCF & 0.0789 & 0.5077 & 0.1206 & 1.1142 & 52.85 & 119.5\\[-0.5ex]
\hline 
\multicolumn{3}{c|}{Mean} & 0.0898 & 0.4671 & 0.1828 & 1.4409 & 103.6 & 208.4\\%[-0.5ex]
%\hline 
\multicolumn{3}{c|}{Min} & -0.0959 & -0.3612 & 0.0423 & 0.3497 & 144.1 & 196.8\\%[-0.5ex]
%\hline 
\multicolumn{3}{c|}{Max} & 0.2164 & 1.2574 & 0.3420 & 2.4948 & 58.04 & 98.41\\%[-0.5ex]
%\hline 
\multicolumn{3}{c|}{Median} & 0.0789 & 0.5077 & 0.2053 & 1.3104 & 160.2 & 158.1\\%[-0.5ex]
\hline 

\end{tabular}
\begin{tablenotes}
\footnotesize
\item Note: Return is the annualized return, displayed in number, not in percentage. Sharpe is the annualized Sharpe ratio. Improvement is defined as that in Table \ref{bigsmall_big}
\end{tablenotes}
\end{threeparttable}

\label{bigsmall_all}
\end{table}

\begin{sidewaystable}[h!]
\centering
\caption{In Sample Performance of Pairs Trading on Intergroup Pairs of Big Banks}

\begin{threeparttable}
\begin{tabular}{c|c|c|c|c|c|c|c|c}
\hline 
\multirow{2}{1.5em}{Pair} & \multirow{2}{4.5em}{Stock \#1} & \multirow{2}{4.5em}{Stock \#2} & \multicolumn{2}{c|}{Model $\textrm{I}$ + Strategy A} & \multicolumn{2}{c|}{Model $\textrm{II}$ + Strategy C} &  \multicolumn{2}{c}{Improvement (in \%)}\tabularnewline
\cline{4-9} \cline{5-9} \cline{6-9} \cline{7-9} \cline{8-9} \cline{9-9} 
 &  &  & {$\,\,\,\,$Return$\,\,\,\,$} & {Sharpe} & {$\,\,\,\,$Return$\,\,\,\,$} & {Sharpe}& {$\,\,\,\,$Return$\,\,\,\,$} & {$\,$Sharpe$\,$} \tabularnewline
\hline \hline 

1 & JPM & BAC & 0.1145 & 0.8864 & 0.1501 & 1.8003 & 31.09 & 103.1\tabularnewline
%\hline 
2 & JPM & WFC & 0.0160 & 0.1461 & 0.0795 & 0.9451 & 396.9 & 546.9\tabularnewline
%\hline 
3 & JPM & C & 0.0664 & 0.5686 & 0.1013 & 1.5193 & 52.56 & 167.2 \tabularnewline
%\hline 
4 & JPM & USB & 0.0186 & 0.2172 & 0.0629 & 1.4293 & 238.2 & 558.1 \tabularnewline
%\hline 
5 & BAC & WFC & 0.0027 & 0.0179 & 0.0568 & 0.4748 & 2004 & 2553 \tabularnewline
%\hline 
6 & BAC & C & 0.0920 & 0.7252 & 0.1193 & 1.5417 & 29.67 & 112.6\tabularnewline
%\hline 
7 & BAC & USB & 0.0603 & 0.3936 & 0.1535 & 1.5144 & 154.6 & 284.8 \tabularnewline
%\hline 
8 & WFC & C & 0.0827 & 0.5918 & 0.1219 & 1.2283 & 47.40 & 107.6\tabularnewline
%\hline 
9 & WFC & USB & 0.0600 & 0.6432 & 0.0739 & 0.9603 & 23.17 & 49.30 \tabularnewline
%\hline 
10 & C & USB & 0.1146 & 0.8553 & 0.1695 & 1.7648 & 47.91 & 106.3\tabularnewline
\hline 
\multicolumn{3}{c|}{Mean} & 0.0628 & 0.5045 & 0.1089 & 1.3178 & 73.42 & 161.2 \tabularnewline
%\hline 
\multicolumn{3}{c|}{Min} & 0.0027 & 0.0179 & 0.0568 & 0.4748 & 2004 & 2553\tabularnewline
%\hline 
\multicolumn{3}{c|}{Max} & 0.1146 & 0.8864 & 0.1695 & 1.8003 & 47.91 & 103.1\tabularnewline
%\hline 
\multicolumn{3}{c|}{Median} & 0.0634 & 0.5802 & 0.1103 & 1.4719 & 74.11 & 153.7\tabularnewline
\hline 
\end{tabular}
\begin{tablenotes}
\footnotesize
\item Note: The data is from 01/10/2012 to 01/01/2018. Return is the annualized return, displayed in number, not in percentage. Sharpe is the annualized Sharpe ratio. Improvement is defined as that in Table \ref{bigsmall_big}.
\end{tablenotes}
\end{threeparttable}

\label{bigsmallout_big_in}
\end{sidewaystable}

\begin{sidewaystable}[h!]
\centering
\caption{Out of Sample Performance of Pairs Trading on Intergroup Pairs of Big Banks}

\begin{threeparttable}
\begin{tabular}{c|c|c|c|c|c|c|c|c}
\hline 
\multirow{2}{1.5em}{Pair} & \multirow{2}{4.5em}{Stock \#1} & \multirow{2}{4.5em}{Stock \#2} & \multicolumn{2}{c|}{Model $\textrm{I}$ + Strategy A} & \multicolumn{2}{c|}{Model $\textrm{II}$ + Strategy C} &  \multicolumn{2}{c}{Improvement (in \%)}\tabularnewline
\cline{4-9} \cline{5-9} \cline{6-9} \cline{7-9} \cline{8-9} \cline{9-9} 
 &  &  & {$\,\,\,\,$Return$\,\,\,\,$} & {Sharpe} & {$\,\,\,\,$Return$\,\,\,\,$} & {Sharpe}& {$\,\,\,\,$Return$\,\,\,\,$} & {$\,$Sharpe$\,$} \tabularnewline
\hline \hline 

1 & JPM & BAC & -0.0503 & -0.4730 & -0.0500 & -0.4760 & 0.5964 & -0.6342\tabularnewline
%\hline 
2 & JPM & WFC & -0.0809 & -0.5693 & -0.0361 & -0.3281 & 55.38 & 42.37\tabularnewline
%\hline 
3 & JPM & C & -0.0841 & -0.6845 & 0.0299 & 0.3228 & 135.6 & 147.2 \tabularnewline
%\hline 
4 & JPM & USB & 0.0867 & 0.9267 & 0.1297 & 1.6816 & 49.60 & 81.46 \tabularnewline
%\hline 
5 & BAC & WFC & 0.0364 & 0.4593 & 0.0464 & 0.4636 & 27.47 & 0.9362 \tabularnewline
%\hline 
6 & BAC & C & -0.0512 & -0.3766 & 0.0149 & 0.2612 & 129.1 & 169.4\tabularnewline
%\hline 
7 & BAC & USB & -0.0037 & -0.0252 & 0.0587 & 0.5169 & 1686 & 2151 \tabularnewline
%\hline 
8 & WFC & C & -0.0586 & -0.3472 & 0.0698 & 0.7619 & 219.1 & 319.5\tabularnewline
%\hline 
9 & WFC & USB & -0.1029 & -0.6961 & 0.0269 & 0.3591 & 126.4 & 151.6 \tabularnewline
%\hline 
10 & C & USB & -0.0486 & -0.2948 & 0.0942 & 0.7796 & 293.8 & 364.5\tabularnewline
\hline 
\multicolumn{3}{c|}{Mean} & -0.0357 & -0.2081 & 0.0384 & 0.4343 & 207.6 & 308.7 \tabularnewline
%\hline 
\multicolumn{3}{c|}{Min} & -0.1029 & -0.6961 & 0.0500 & -0.4760 & 51.41 & 31.62\tabularnewline
%\hline 
\multicolumn{3}{c|}{Max} & 0.0867 & 0.9267 & 0.1297 & 1.6816 & 49.60 & 81.46\tabularnewline
%\hline 
\multicolumn{3}{c|}{Median} & -0.0508 & -0.3619 & 0.0382 & 0.4114 & 175
2& 213.7\tabularnewline
\hline 
\end{tabular}
\begin{tablenotes}
\footnotesize
\item Note: The data is from 01/01/2018 to 01/12/2019. Return is the annualized return, displayed in number, not in percentage. Sharpe is the annualized Sharpe ratio. Improvement is defined as that in Table \ref{bigsmall_big}.
\end{tablenotes}
\end{threeparttable}

\label{bigsmallout_big_out}
\end{sidewaystable}

\begin{sidewaystable}[h!]
\centering
\caption{In Sample Performance of Pairs Trading on Intergroup Pairs of Small Banks}

\begin{threeparttable}
\begin{tabular}{c|c|c|c|c|c|c|c|c}
\hline 
\multirow{2}{1.5em}{Pair} & \multirow{2}{4.5em}{Stock \#1} & \multirow{2}{4.5em}{Stock \#2} & \multicolumn{2}{c|}{Model $\textrm{I}$ + Strategy A} & \multicolumn{2}{c|}{Model $\textrm{II}$ + Strategy C} &  \multicolumn{2}{c}{Improvement (in \%)}\tabularnewline
\cline{4-9} \cline{5-9} \cline{6-9} \cline{7-9} \cline{8-9} \cline{9-9} 
 &  &  & {$\,\,\,\,$Return$\,\,\,\,$} & {Sharpe} & {$\,\,\,\,$Return$\,\,\,\,$} & {Sharpe}& {$\,\,\,\,$Return$\,\,\,\,$} & {$\,$Sharpe$\,$} \tabularnewline
\hline \hline 

1 & CPF & BANC & 0.2713 & 0.9758 & 0.3513 & 2.0574 & 29.56 & 110.8 \tabularnewline
%\hline 
2 & CPF & CUBI & 0.1226 & 0.4404 & 0.4457 & 1.9114 & 263.5 & 334.0 \tabularnewline
%\hline 
3 & CPF & NBHC & 0.1905 & 0.9823 & 0.2559 & 1.7188 & 34.33 & 74.98  \tabularnewline
%\hline 
4 & CPF & FCF & 0.1855 & 1.2385 & 0.2453 & 2.5505 & 32.24 & 105.9\tabularnewline
%\hline 
5 & BANC & CUBI & 0.2500 & 0.6928 & 0.4076 & 1.9505 & 63.04 & 181.5\tabularnewline
%\hline 
6 & BANC & NBHC & 0.2406 & 0.8926 & 0.1699 & 1.4127 & -29.38 & 58.27\tabularnewline
%\hline 
7 & BANC & FCF & 0.2056 & 0.7819 & 0.3308 & 1.8279 & 60.89 & 133.8 \tabularnewline
%\hline 
8 & CUBI & NBHC & 0.1130 & 0.3808 & 0.2164 & 1.8059 & 91.50 & 374.2\tabularnewline
%\hline 
9 & CUBI & FCF & 0.1125 & 0.4133 & 0.1886 & 1.1579 & 67.64 & 180.2\tabularnewline
%\hline 
10 & NBHC & FCF & 0.1026 & 0.5723 & 0.2523 & 1.8035 & 145.9 & 215.1\tabularnewline
\hline 
\multicolumn{3}{c|}{Mean} & 0.1794 & 0.7371 & 0.2864 & 1.8197 & 59.63 & 146.9\tabularnewline
%\hline 
\multicolumn{3}{c|}{Min} & 0.1026 & 0.3808 & 0.1699 & 1.1579 & 65.59 & 204.1 \tabularnewline
%\hline 
\multicolumn{3}{c|}{Max} & 0.2713 & 1.2385 & 0.4457 & 2.5505 & 64.28 & 105.9 \tabularnewline
%\hline 
\multicolumn{3}{c|}{Median} & 0.1880 & 0.7374 & 0.2541 & 1.8169 & 35.16 & 146.4 \tabularnewline
\hline 

\end{tabular}
\begin{tablenotes}
\footnotesize
\item Note: The data is from 01/10/2012 to 01/01/2018. Return is the annualized return, displayed in number, not in percentage. Sharpe is the annualized Sharpe ratio. Improvement is defined as that in Table \ref{bigsmall_big}.
\end{tablenotes}
\end{threeparttable}

\label{bigsmallout_small_in}
\end{sidewaystable}

\begin{sidewaystable}[h!]
\centering
\caption{Out of Sample Performance of Pairs Trading on Intergroup Pairs of Small Banks}

\begin{threeparttable}
\begin{tabular}{c|c|c|c|c|c|c|c|c}
\hline 
\multirow{2}{1.5em}{Pair} & \multirow{2}{4.5em}{Stock \#1} & \multirow{2}{4.5em}{Stock \#2} & \multicolumn{2}{c|}{Model $\textrm{I}$ + Strategy A} & \multicolumn{2}{c|}{Model $\textrm{II}$ + Strategy C} &  \multicolumn{2}{c}{Improvement (in \%)}\tabularnewline
\cline{4-9} \cline{5-9} \cline{6-9} \cline{7-9} \cline{8-9} \cline{9-9} 
 &  &  & {$\,\,\,\,$Return$\,\,\,\,$} & {Sharpe} & {$\,\,\,\,$Return$\,\,\,\,$} & {Sharpe}& {$\,\,\,\,$Return$\,\,\,\,$} & {$\,$Sharpe$\,$} \tabularnewline
\hline \hline 

1 & CPF & BANC & 0.1856 & 0.7541 & 0.1649 & 0.8297 & -11.15 & 10.03 \tabularnewline
%\hline 
2 & CPF & CUBI & -0.0924 & -0.3528 & 0.2424 & 1.8467 & 362.3 & 623.4 \tabularnewline
%\hline 
3 & CPF & NBHC & -0.0769 & -0.3944 & 0.1621 & 1.0216 & 310.8 & 359.0  \tabularnewline
%\hline 
4 & CPF & FCF & -0.0373 & -0.1906 & 0.2094 & 1.4249 & 661.4 & 847.6\tabularnewline
%\hline 
5 & BANC & CUBI & 0.1266 & 0.7454 & 0.4109 & 2.5902 & 224.6 & 247.5\tabularnewline
%\hline 
6 & BANC & NBHC & -0.1577 & -0.6720 & -0.0797 & -0.3926 & 49.46 & 41.58\tabularnewline
%\hline 
7 & BANC & FCF & 0.0107 & 0.0821 & 0.1601 & 1.3930 & 1396 & 1596 \tabularnewline
%\hline 
8 & CUBI & NBHC & -0.1475 & -0.5514 & 0 & - & 100 & 100\tabularnewline
%\hline 
9 & CUBI & FCF & -0.1137 & -0.4079 & 0 & - & 100 & 100\tabularnewline
%\hline 
10 & NBHC & FCF & -0.0578 & -0.3088 & 0.1520 & 1.0421 & 363.0 & 437.4\tabularnewline
\hline 
\multicolumn{3}{c|}{Mean} & -0.0360 & -0.1296 & 0.1422 & 0.9756 & 494.6 & 852.6\tabularnewline
%\hline 
\multicolumn{3}{c|}{Min} & -0.1577 & -0.6720 & -0.0797 & -0.3926 & 49.46 & 41.58 \tabularnewline
%\hline 
\multicolumn{3}{c|}{Max} & 0.1856 & 0.7541 & 0.4109 & 2.5902 & 121.4 & 243.5 \tabularnewline
%\hline 
\multicolumn{3}{c|}{Median} & -0.0674 & -0.3308 & 0.1611 & 1.0319 & 339.2 & 411.9\tabularnewline
\hline 

\end{tabular}
\begin{tablenotes}
\footnotesize
\item Note: The data is from 01/01/2018 to 01/12/2019. Return is the annualized return, displayed in number, not in percentage. Sharpe is the annualized Sharpe ratio. Improvement is defined as that in Table \ref{bigsmall_big}. The returns for CUBI/NBHC and CUBI/FCF are 0 because no trading is opened for these two pairs during the out-of-sample period, and the Sharpe ratios are undefined.
\end{tablenotes}
\end{threeparttable}

\label{bigsmallout_small_out}
\end{sidewaystable}

\begin{table}[h!]
\centering
\caption{In Sample Performance of Pairs Trading on Intragroup Pairs}

\begin{threeparttable}
\begin{tabular}{c|c|c|c|c|c|c|c|c}
\hline 
\multirow{2}{1.5em}{Pair} & \multirow{2}{3.2em}{Stock \#1} & \multirow{2}{3.2em}{Stock \#2} & \multicolumn{2}{c|}{Model $\textrm{I}$ + Strategy A} & \multicolumn{2}{c|}{Model $\textrm{II}$ + Strategy C} & \multicolumn{2}{c}{Improvement (in \%)}\tabularnewline
\cline{4-9} \cline{5-9} \cline{6-9} \cline{7-9} \cline{8-9} \cline{9-9} 
 &  &  & {$\,\,\,\,$Return$\,\,\,\,$} & {Sharpe} & {$\,\,\,\,$Return$\,\,\,\,$} & {Sharpe}& {$\,\,\,\,$Return$\,\,\,\,$} & {$\,\,$Sharpe$\,\,$} \tabularnewline
 \hline\hline

1 & JPM & CPF & 0.1668 & 0.9415 & 0.2866 & 3.0567 & 71.82 & 224.7\\[-0.5ex]
%\hline 
2 & JPM & BANC & 0.2067 & 0.7134 & 0.2581 & 1.5501 & 24.87 & 117.3 \\[-0.5ex]
%\hline 
3 & JPM & CUBI & 0.0649 & 0.9832 & 0.2576 & 1.6633 & 296.9 & 69.17\\[-0.5ex]
%\hline 
4 & JPM & NBHC & 0.1505 & 0.8387 & 0.2735 & 2.2745 & 81.73 & 171.2\\[-0.5ex]
%\hline 
5 & JPM & FCF & 0.2083 & 1.3273 & 0.3281 & 2.9235 & 57.51 & 120.3\\[-0.5ex]
%\hline 
6 & BAC & CPF & 0.1572 & 0.7484 & 0.2099 & 1.7310 & 33.52 & 131.3 \\[-0.5ex]
%\hline 
7 & BAC & BANC & 0.2361 & 0.7452 & 0.1708 & 1.0044 & -27.66 & 34.78\\[-0.5ex]
%\hline 
8 & BAC & CUBI & 0.0789 & 0.2755 & 0.1669 & 1.4519 & 111.5 & 427.0 \\[-0.5ex]
%\hline 
9 & BAC & NBHC & 0.2608 & 1.2323 & 0.3354 & 2.5663 & 28.60 & 108.3 \\[-0.5ex]
%\hline 
10 & BAC & FCF & 0.1918 & 1.0401 & 0.2653 & 2.3337 & 38.32 & 124.4\\[-0.5ex]
%\hline 
11 & WFC & CPF & 0.0376 & 0.1924 & 0.0988 & 0.6388 & 162.8 & 232.0\\[-0.5ex]
%\hline 
12 & WFC & BANC & 0.2371 & 0.8323 & 0.2165 & 1.0599 & -8.690 & 27.53\\[-0.5ex]
%\hline 
13 & WFC & CUBI & 0.0729 & 0.2682 & 0.2307 & 1.9597 & 216.5 & 630.7\\[-0.5ex]
%\hline 
14 & WFC & NBHC & 0.0974 & 0.5548 & 0.0917 & 0.6167 & -5.850 & 11.16\\[-0.5ex]
%\hline 
15 & WFC & FCF & 0.0656 & 0.3971 & 0.1413 & 1.1406 & 115.4 & 187.2\\[-0.5ex]
%\hline 
16 & C & CPF & 0.0571 & 0.2873 & 0.1766 & 1.4015 & 206.3 & 387.8\\[-0.5ex]
%\hline 
17 & C & BANC & 0.2454 & 0.8899 & 0.2154 & 1.9512 & -12.22 & 119.3\\[-0.5ex]
%\hline 
18 & C & CUBI & 0.0715 & 0.2696 & 0.1589 & 1.0954 & 122.2 & 306.3\\[-0.5ex]
%\hline 
19 & C & NBHC & 0.1279 & 0.6511 & 0.2125 & 1.5321 & 66.15 & 135.3\\[-0.5ex]
%\hline 
20 & C & FCF & 0.1160 & 0.6154 & 0.1790 & 1.3736 & 54.31 & 123.2\\[-0.5ex]
%\hline 
21 & USB & CPF & 0.0654 & 0.4915 & 0.2126 & 1.9990 & 225.1 & 306.7\\[-0.5ex]
%\hline 
22 & USB & BANC & 0.2164 & 0.7529 & 0.3389 & 1.9118 & 56.61 & 153.9\\[-0.5ex]
%\hline 
23 & USB & CUBI & 0.0565 & 0.2443 & 0.2826 & 1.9450 & 400.2 & 696.2\\[-0.5ex]
%\hline 
24 & USB & NBHC & 0.1340 & 0.9289 & 0.1947 & 1.5321 & 45.30 & 64.94\\[-0.5ex]
%\hline 
25 & USB & FCF & 0.0922 & 0.6221 & 0.2167 & 2.1579 & 135.0 & 246.9\\[-0.5ex]
\hline 
\multicolumn{3}{c|}{Mean} & 0.1366 & 0.6737 & 0.2208 & 1.7148 & 61.61 & 154.5\\%[-0.5ex]
%\hline 
\multicolumn{3}{c|}{Min} & 0.0376 & 0.1924 & 0.0917 & 0.6167 & 143.9 & 220.5\\%[-0.5ex]
%\hline 
\multicolumn{3}{c|}{Max} & 0.2608 & 1.3273 & 0.3389 & 3.0567 & 29.95 & 130.3\\%[-0.5ex]
%\hline 
\multicolumn{3}{c|}{Median} & 0.1279 & 0.7134 & 0.2154 & 1.6633 & 68.41 & 133.2\\%[-0.5ex]
\hline 

\end{tabular}
\begin{tablenotes}
\footnotesize
\item Note: The data is from 01/10/2012 to 01/01/2018. Return is the annualized return, displayed in number, not in percentage. Sharpe is the annualized Sharpe ratio. Improvement is defined as that in Table \ref{bigsmall_big}.
\end{tablenotes}
\end{threeparttable}

\label{bigsmallout_all_in}
\end{table}

\begin{table}[h!]
\centering
\caption{Out of Sample Performance of Pairs Trading on Intragroup Pairs}

\begin{threeparttable}
\begin{tabular}{c|c|c|c|c|c|c|c|c}
\hline 
\multirow{2}{1.5em}{Pair} & \multirow{2}{3.2em}{Stock \#1} & \multirow{2}{3.2em}{Stock \#2} & \multicolumn{2}{c|}{Model $\textrm{I}$ + Strategy A} & \multicolumn{2}{c|}{Model $\textrm{II}$ + Strategy C} & \multicolumn{2}{c}{Improvement (in \%)}\tabularnewline
\cline{4-9} \cline{5-9} \cline{6-9} \cline{7-9} \cline{8-9} \cline{9-9} 
 &  &  & {$\,\,\,\,$Return$\,\,\,\,$} & {Sharpe} & {$\,\,\,\,$Return$\,\,\,\,$} & {Sharpe}& {$\,\,\,\,$Return$\,\,\,\,$} & {$\,$Sharpe$\,$} \tabularnewline
 \hline\hline

1 & JPM & CPF & 0.1514 & 0.8997 & 0.2731 & 2.3058 & 80.38 & 156.3\\[-0.5ex]
%\hline 
2 & JPM & BANC & 0.2190 & 0.9752 & 0.2023 & 1.1630 & -7.626 & 19.26 \\[-0.5ex]
%\hline 
3 & JPM & CUBI & 0.0965 & 1.1227 & 0.1610 & 1.0135 & 66.84 & -9.727\\[-0.5ex]
%\hline 
4 & JPM & NBHC & 0.0303 & 0.1492 & 0.1799 & 1.8165 & 493.7 & 1117\\[-0.5ex]
%\hline 
5 & JPM & FCF & 0.0878 & 0.4209 & 0.1682 & 1.0338 & 91.57 & 145.6\\[-0.5ex]
%\hline 
6 & BAC & CPF & 0.0379 & 0.1702 & 0.1592 & 1.3579 & 320.1 & 697.8 \\[-0.5ex]
%\hline 
7 & BAC & BANC & 0.1763 & 0.6913 & 0.1693 & 0.8830 & -3.971 & 27.73\\[-0.5ex]
%\hline 
8 & BAC & CUBI & 0.0926 & 0.3435 & 0.1014 & 0.4298 & 9.503 & 25.12 \\[-0.5ex]
%\hline 
9 & BAC & NBHC & -0.0212 & -0.0999 & 0.0144 & 0.7148 & 167.9 & 815.5 \\[-0.5ex]
%\hline 
10 & BAC & FCF & 0.0196 & 0.0899 & 0.1117 & 0.8152 & 469.9 & 8.6.8\\[-0.5ex]
%\hline 
11 & WFC & CPF & -0.0625 & -0.2981 & -0.0061 & 0.6388 & 90.24 & 314.3\\[-0.5ex]
%\hline 
12 & WFC & BANC & 0.0583 & 0.2249 & 0.1282 & 0.6058 & 119.9 & 169.4\\[-0.5ex]
%\hline 
13 & WFC & CUBI & -0.0181 & -0.0652 & 0.2826 & 1.5870 & 1661 & 2534\\[-0.5ex]
%\hline 
14 & WFC & NBHC & -0.1181 & -0.5631 & 0.0447 & 0.2594 & 137.8 & 146.1\\[-0.5ex]
%\hline 
15 & WFC & FCF & -0.0821 & -0.3725 & 0.1225 & 0.8413 & 249.2 & 325.9\\[-0.5ex]
%\hline 
16 & C & CPF & -0.0072 & -0.0314 & 0.1433 & 1.1894 & 2090 & 3888\\[-0.5ex]
%\hline 
17 & C & BANC & 0.1238 & 0.4691 & 0.0839 & 0.6480 & -32.23 & 38.13\\[-0.5ex]
%\hline 
18 & C & CUBI & 0.0459 & 0.1692 & 0.2568 & 1.2778 & 459.5 & 655.2\\[-0.5ex]
%\hline 
19 & C & NBHC & -0.0648 & -0.2911 & 0.2108 & 2.1138 & 425.3 & 826.1\\[-0.5ex]
%\hline 
20 & C & FCF & -0.0265 & -0.1143 & 0.2174 & 1.2651 & 920.4 & 1207\\[-0.5ex]
%\hline 
21 & USB & CPF & 0.2108 & 2.2429 & 0.2652 & 2.4946 & 25.81 & 11.22\\[-0.5ex]
%\hline 
22 & USB & BANC & 0.1951 & 0.8939 & 0.1909 & 1.3332 & -2.153 & 49.14\\[-0.5ex]
%\hline 
23 & USB & CUBI & 0.1516 & 0.7685 & 0.2356 & 1.5712 & 55.41 & 104.5\\[-0.5ex]
%\hline 
24 & USB & NBHC & -0.0242 & -0.1258 & 0.1514 & 0.9637 & 725.6 & 866.1\\[-0.5ex]
%\hline 
25 & USB & FCF & 0.0037 & 0.0192 & 0.1979 & 1.2151 & 5249 & 6229\\[-0.5ex]
\hline 
\multicolumn{3}{c|}{Mean} & 0.0510 & 0.3076 & 0.1626 & 1.1815 & 218.6 & 284.2\\%[-0.5ex]
%\hline 
\multicolumn{3}{c|}{Min} & -0.1181 & -0.5631 & -0.0061 & 0.2594 & 94.84 & 146.4\\%[-0.5ex]
%\hline 
\multicolumn{3}{c|}{Max} & 0.2190 & 2.2429 & 0.2826 & 2.4946 & 29.04 & 11.22\\%[-0.5ex]
%\hline 
\multicolumn{3}{c|}{Median} & 0.0379 & 0.1692 & 0.1682 & 1.1630 & 343.8 & 587.4\\%[-0.5ex]
\hline 

\end{tabular}
\begin{tablenotes}
\footnotesize
\item Note: The data is from 01/01/2018 to 01/12/2019. Return is the annualized return, displayed in number, not in percentage. Sharpe is the annualized Sharpe ratio. Improvement is defined as that in Table \ref{bigsmall_big}.
\end{tablenotes}
\end{threeparttable}

\label{bigsmallout_all_out}
\end{table}
\clearpage

%\section{Figures}
\begin{figure}[h!]
\captionsetup{belowskip=12pt,aboveskip=4pt}
\caption{Performance of Strategy A, B and C, based on Model 1}
\begin{subfigure}{.5\textwidth}
  \centering
  \caption{Return of Strategy A, Model 1}
  \includegraphics[scale=0.245]{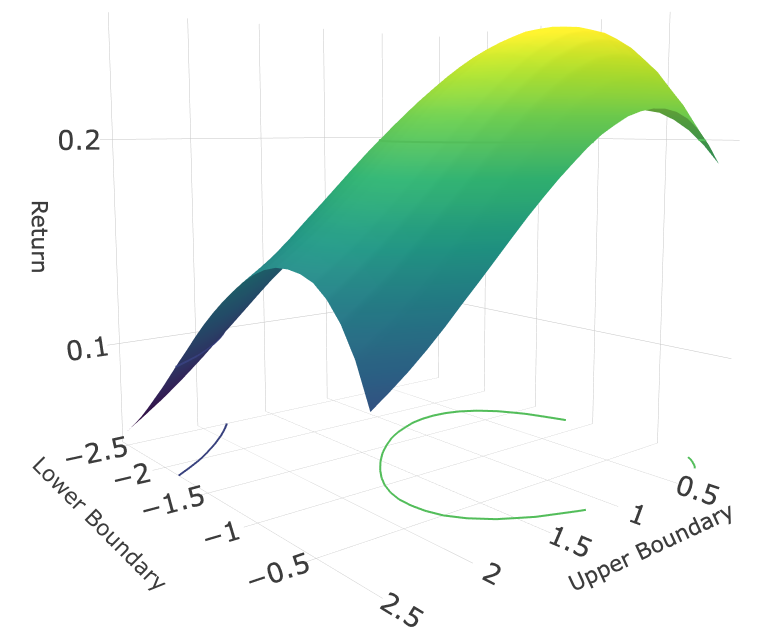}  
  
  \label{fig:M1_strA_tr}
\end{subfigure}
\begin{subfigure}{.5\textwidth}
  \centering
  \caption{Sharpe Ratio of Strategy A, Model 1}
  \includegraphics[scale=0.245]{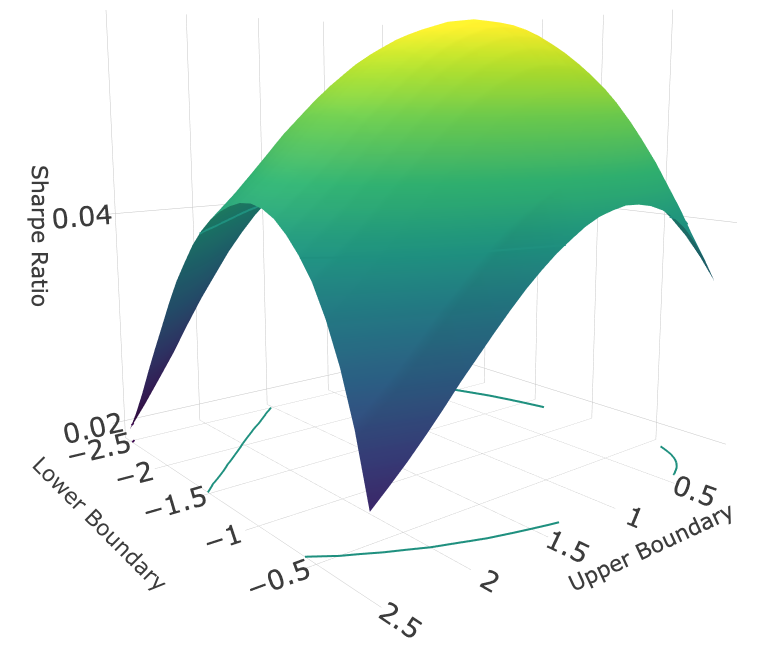}  
  
  \label{fig:M1_strA_sharpe}
\end{subfigure}

%\newline

\begin{subfigure}{.5\textwidth}
  \centering
  \caption{Return of Strategy B, Model 1}
  \includegraphics[scale=0.245]{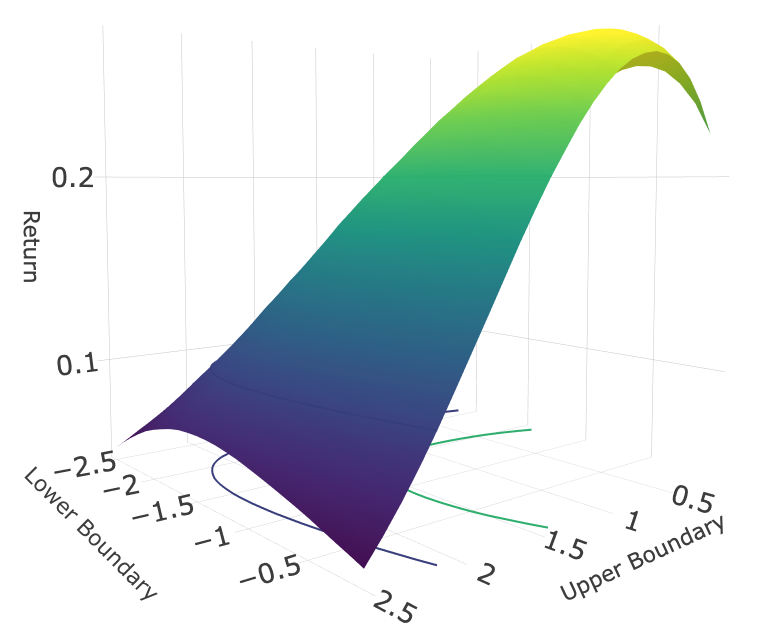}  
  
  \label{fig:M1_strB_tr}
\end{subfigure}
\begin{subfigure}{.5\textwidth}
  \centering
  \caption{Sharpe Ratio of Strategy B, Model 1}
  \includegraphics[scale=0.245]{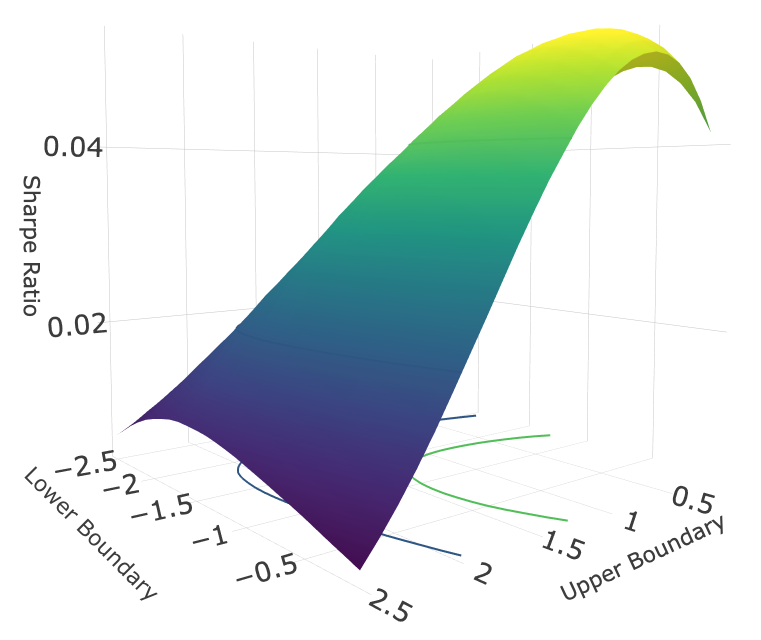}  
  
  \label{fig:M1_strB_sharpe}
\end{subfigure}

%\newline

\begin{subfigure}{.5\textwidth}
  \centering
  \caption{Return of Strategy C, Model 1}
  \includegraphics[scale=0.245]{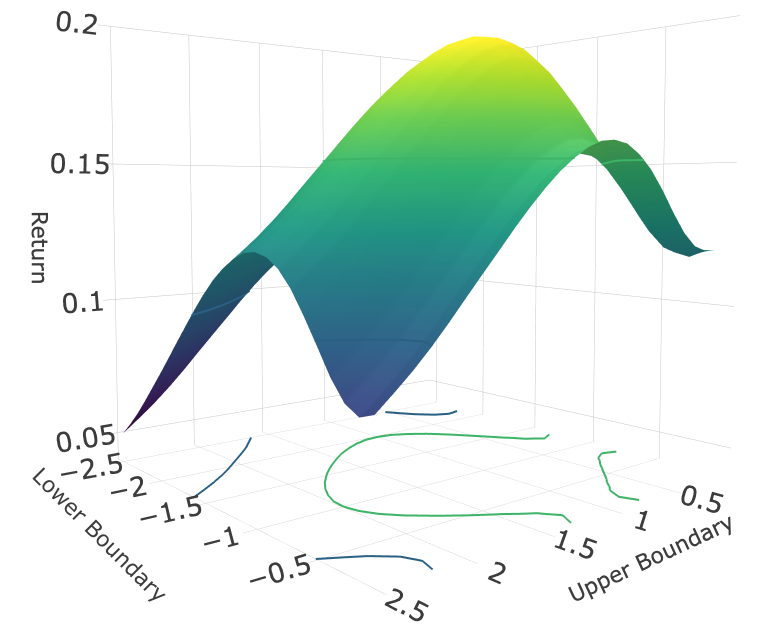}  
  
  \label{fig:M1_strC_tr}
\end{subfigure}
\begin{subfigure}{.5\textwidth}
  \centering
  \caption{Sharpe Ratio of Strategy C, Model 1}
  \includegraphics[scale=0.245]{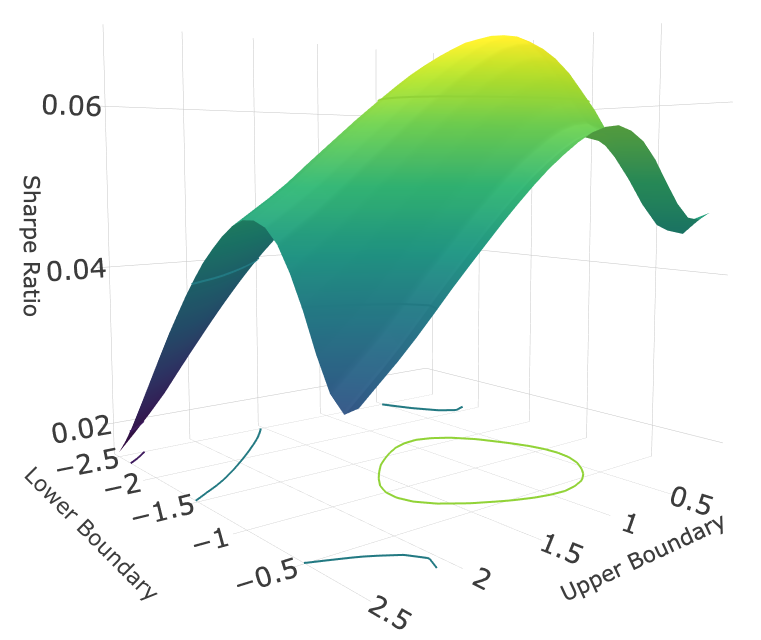}  
  
  \label{fig:M1_strC_sharpe}
\end{subfigure}

\label{fig:model1}
\end{figure}

\begin{figure}[h!]
\captionsetup{belowskip=12pt,aboveskip=4pt}
\caption{Performance of Strategy A, B and C, based on Model 2}
\begin{subfigure}{.5\textwidth}
  \centering
  \caption{Return of Strategy A, Model 2}
  \includegraphics[scale=0.245]{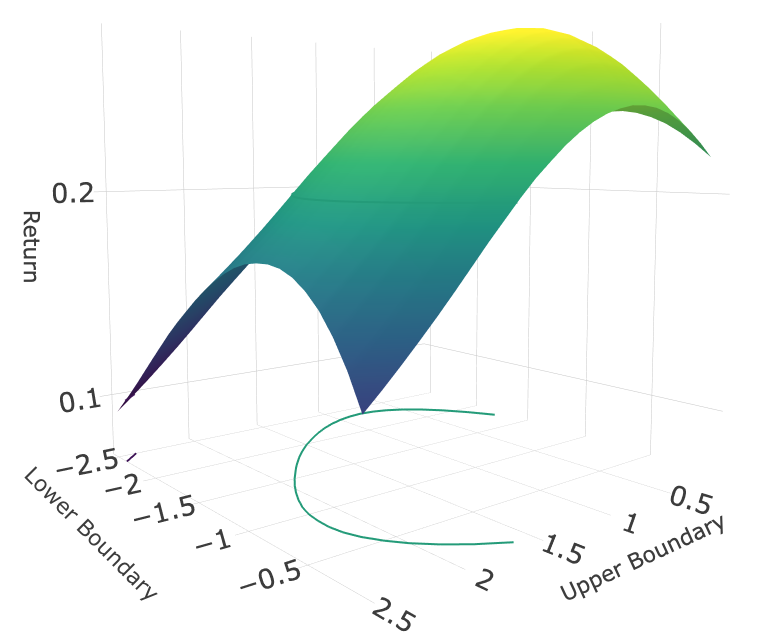}  
  
  \label{fig:M2_strA_tr}
\end{subfigure}
\begin{subfigure}{.5\textwidth}
  \centering
  \caption{Sharpe Ratio of Strategy A, Model 2}
  \includegraphics[scale=0.245]{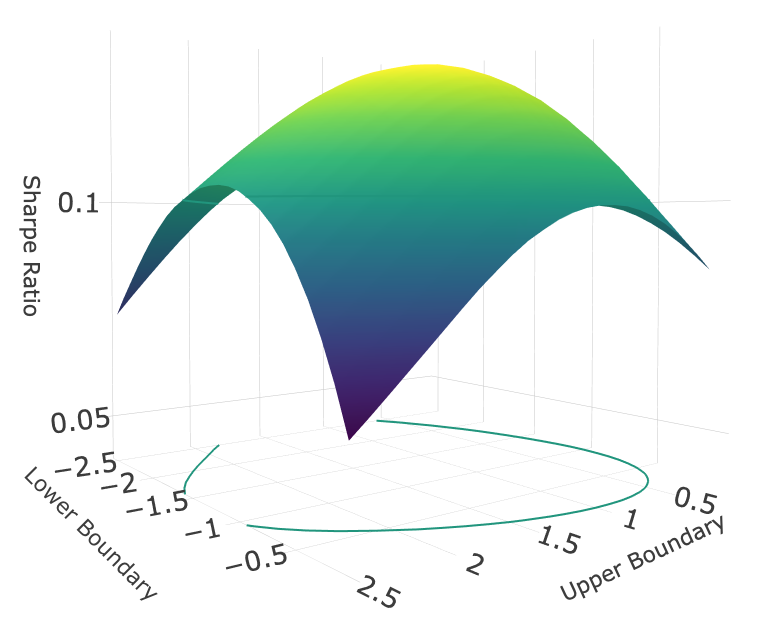}  
  
  \label{fig:M2_strA_sharpe}
\end{subfigure}

%\newline

\begin{subfigure}{.5\textwidth}
  \centering
  \caption{Return of Strategy B, Model 2}
  \includegraphics[scale=0.245]{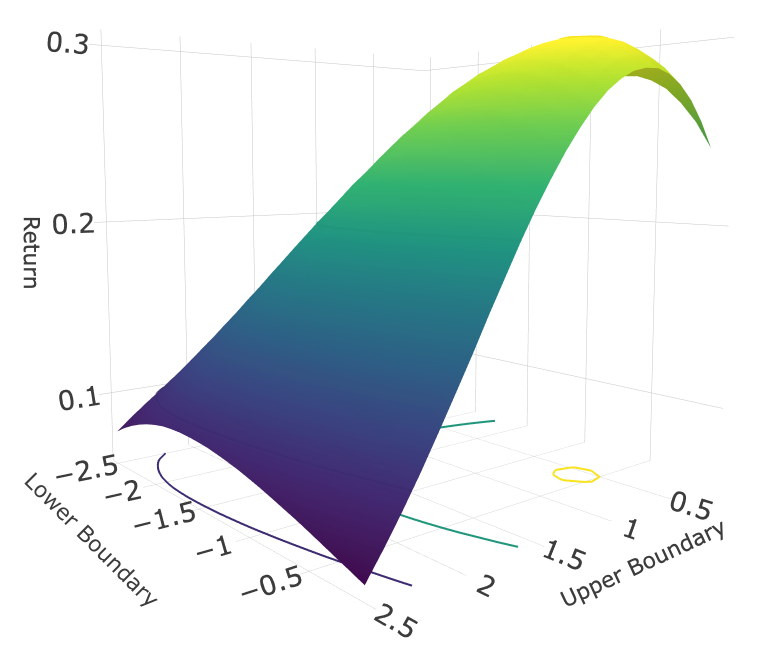}  
  
  \label{fig:M2_strB_tr}
\end{subfigure}
\begin{subfigure}{.5\textwidth}
  \centering
  \caption{Sharpe Ratio of Strategy B, Model 2}
  \includegraphics[scale=0.245]{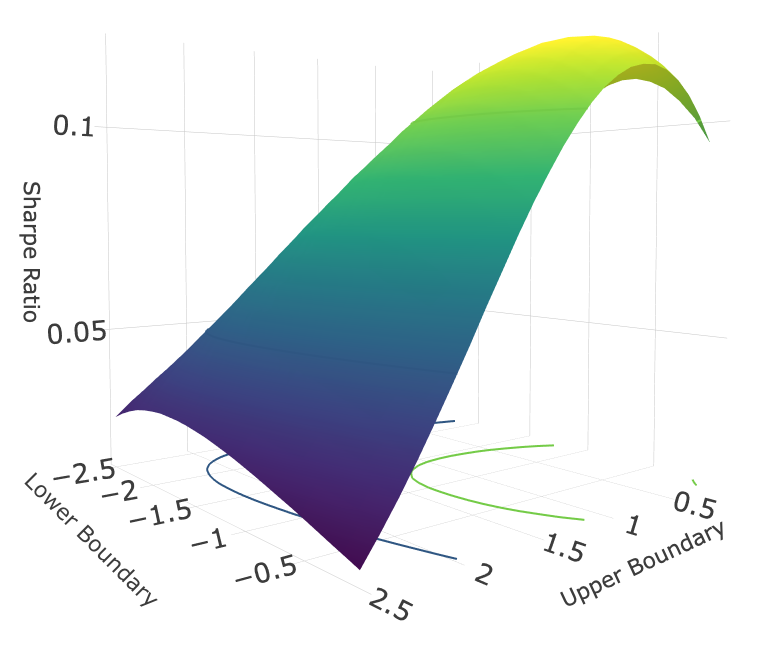}  
  
  \label{fig:M2_strB_sharpe}
\end{subfigure}

%\newline

\begin{subfigure}{.5\textwidth}
  \centering
  \caption{Return of Strategy C, Model 2}
  \includegraphics[scale=0.245]{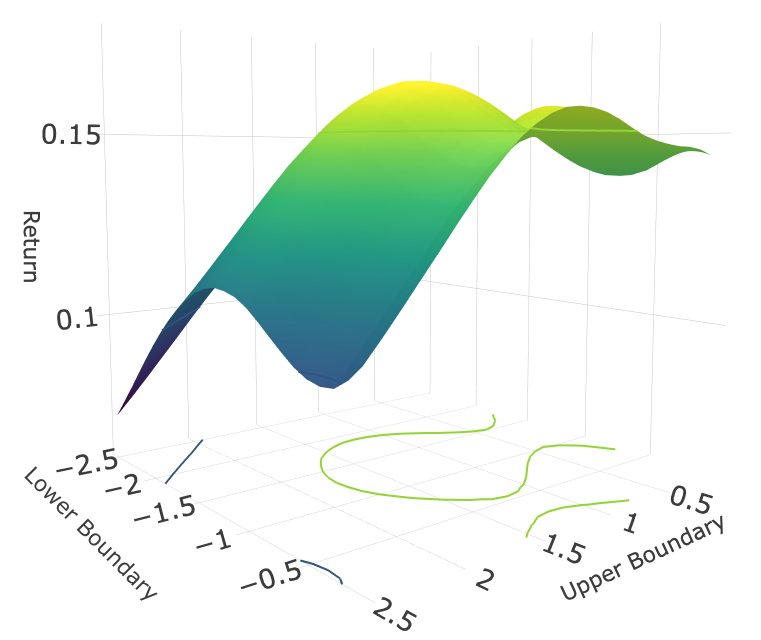}  
  
  \label{fig:M2_strC_tr}
\end{subfigure}
\begin{subfigure}{.5\textwidth}
  \centering
  \caption{Sharpe Ratio of Strategy C, Model 2}
  \includegraphics[scale=0.245]{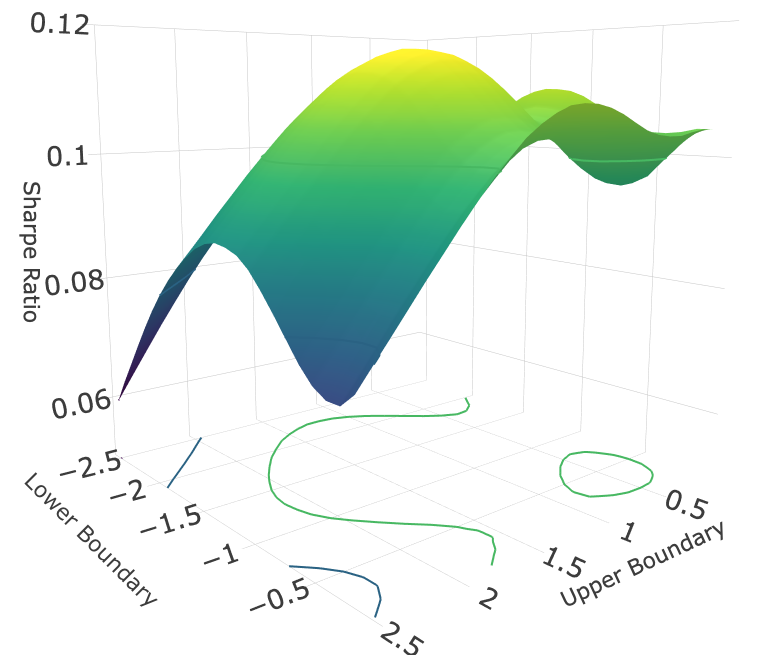}  
  
  \label{fig:M2_strC_sharpe}
\end{subfigure}

\label{fig:model2}
\end{figure}

\begin{figure}[h!]
\captionsetup{belowskip=12pt,aboveskip=4pt}
\caption{Performance of Strategy A, B and C, based on Model 3}
\begin{subfigure}{.5\textwidth}
  \centering
  \caption{Return of Strategy A, Model 3}
  \includegraphics[scale=0.245]{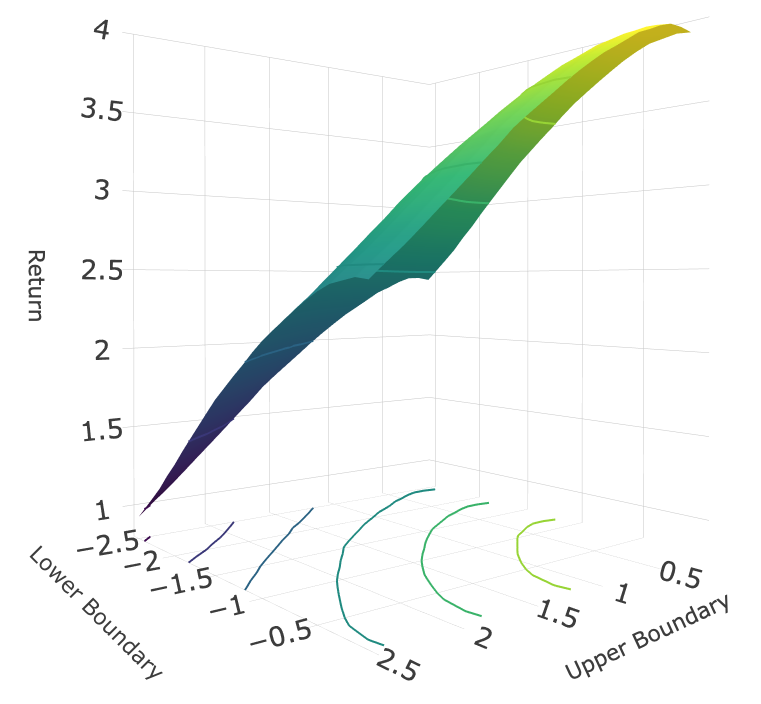}  
  
  \label{fig:M3_strA_tr}
\end{subfigure}
\begin{subfigure}{.5\textwidth}
  \centering
  \caption{Sharpe Ratio of Strategy A, Model 3}
  \includegraphics[scale=0.245]{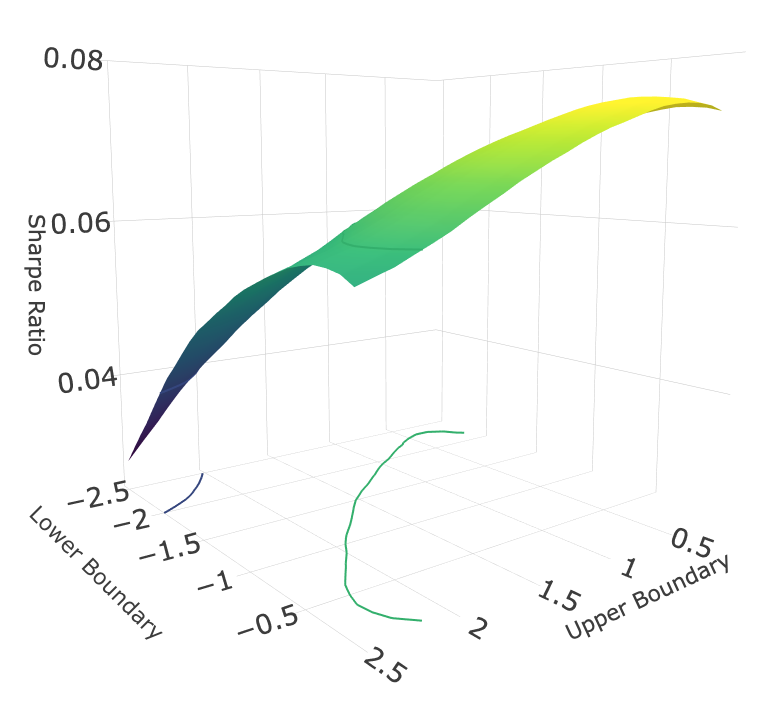}  
  
  \label{fig:M3_strA_sharpe}
\end{subfigure}

%\newline

\begin{subfigure}{.5\textwidth}
  \centering
  \caption{Return of Strategy B, Model 3}
  \includegraphics[scale=0.245]{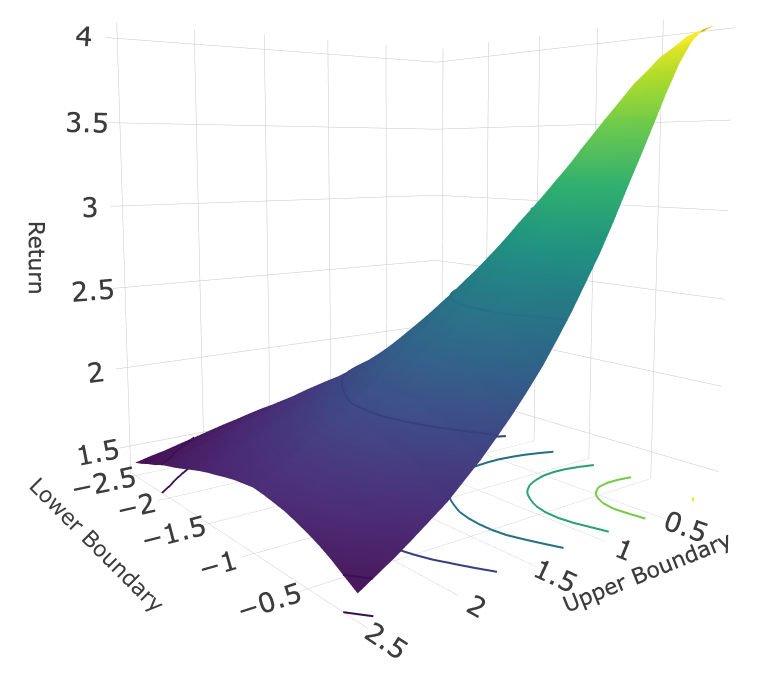}  
  
  \label{fig:M3_strB_tr}
\end{subfigure}
\begin{subfigure}{.5\textwidth}
  \centering
  \caption{Sharpe Ratio of Strategy B, Model 3}
  \includegraphics[scale=0.245]{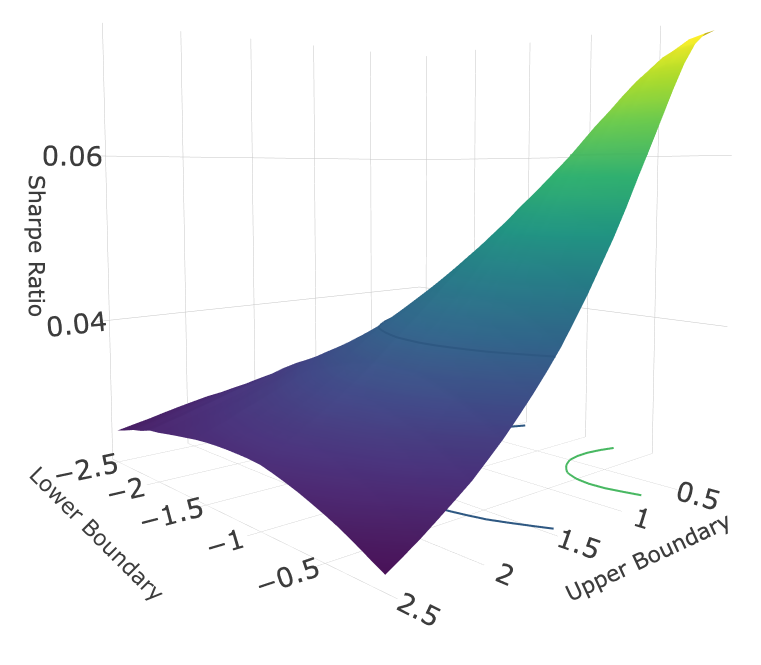}  
  
  \label{fig:M3_strB_sharpe}
\end{subfigure}

%\newline

\begin{subfigure}{.5\textwidth}
  \centering
  \caption{Return of Strategy C, Model 3}
  \includegraphics[scale=0.245]{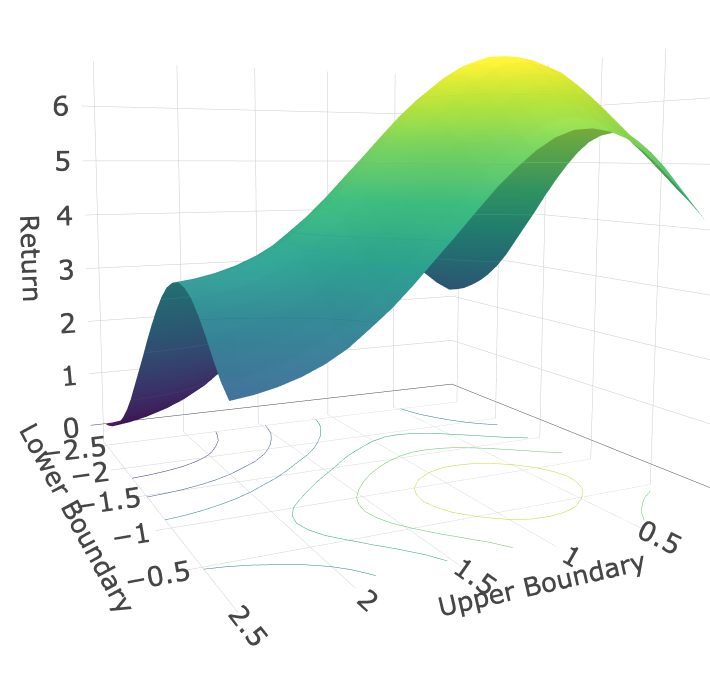}  
  
  \label{fig:M3_strC_tr}
\end{subfigure}
\begin{subfigure}{.5\textwidth}
  \centering
  \caption{Sharpe Ratio of Strategy C, Model 3}
  \includegraphics[scale=0.245]{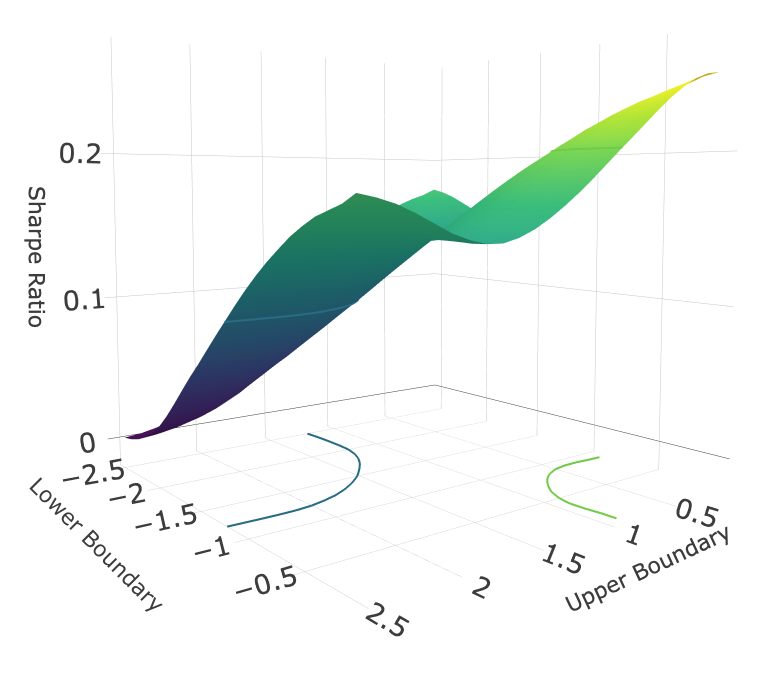}  
  
  \label{fig:M3_strC_sharpe}
\end{subfigure}

\label{fig:model3}
\end{figure}

\begin{figure}[h!]
\captionsetup{belowskip=12pt,aboveskip=4pt}
\caption{Performance of Strategy A, B and C, based on Model 4}
\begin{subfigure}{.5\textwidth}
  \centering
  \caption{Return of Strategy A, Model 4}
  \includegraphics[scale=0.245]{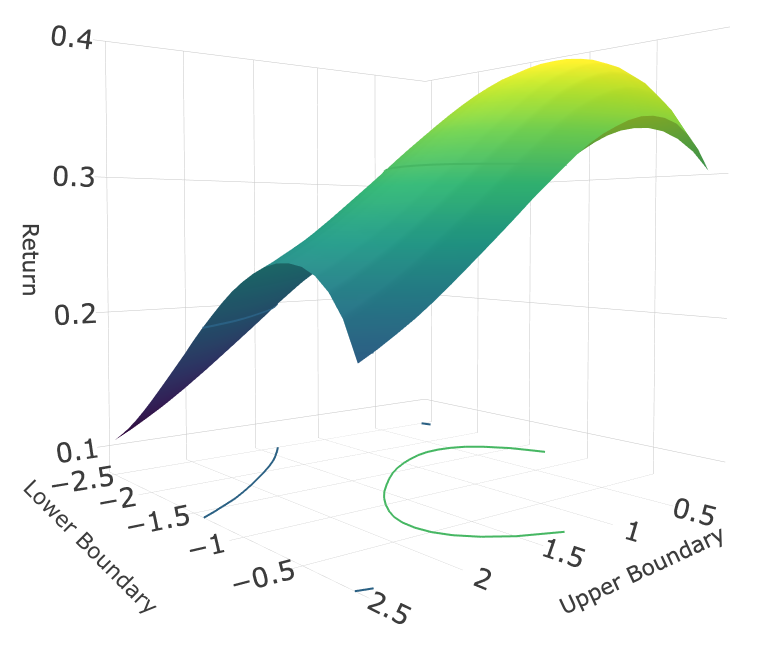}  
  
  \label{fig:M4_strA_tr}
\end{subfigure}
\begin{subfigure}{.5\textwidth}
  \centering
  \caption{Sharpe Ratio of Strategy A, Model 4}
  \includegraphics[scale=0.245]{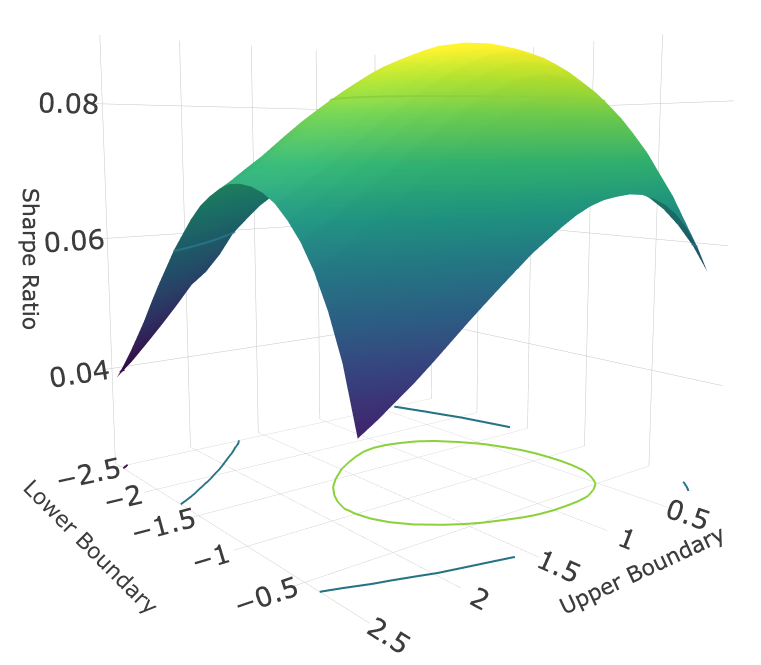}  
  
  \label{fig:M4_strA_sharpe}
\end{subfigure}

%\newline

\begin{subfigure}{.5\textwidth}
  \centering
  \caption{Return of Strategy B, Model 4}
  \includegraphics[scale=0.245]{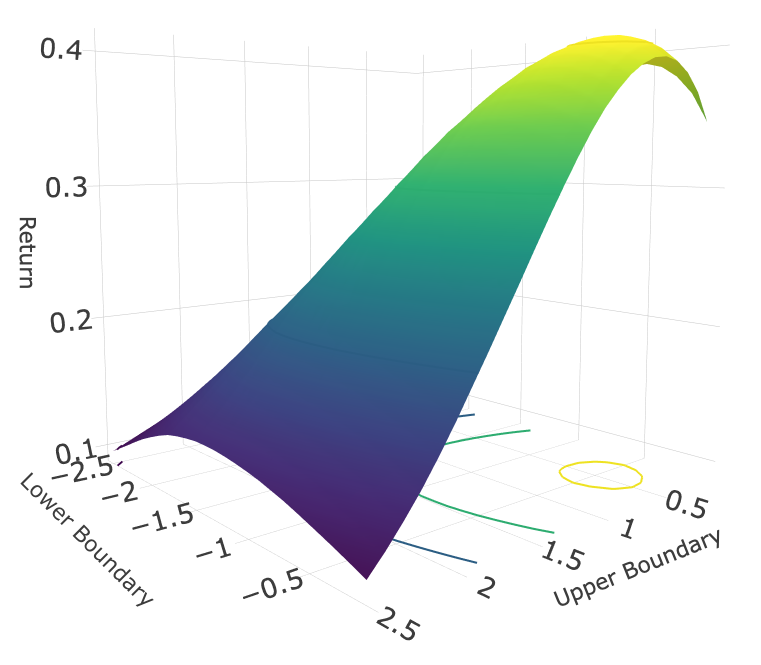}  
  
  \label{fig:M4_strB_tr}
\end{subfigure}
\begin{subfigure}{.5\textwidth}
  \centering
  \caption{Sharpe Ratio of Strategy B, Model 4}
  \includegraphics[scale=0.245]{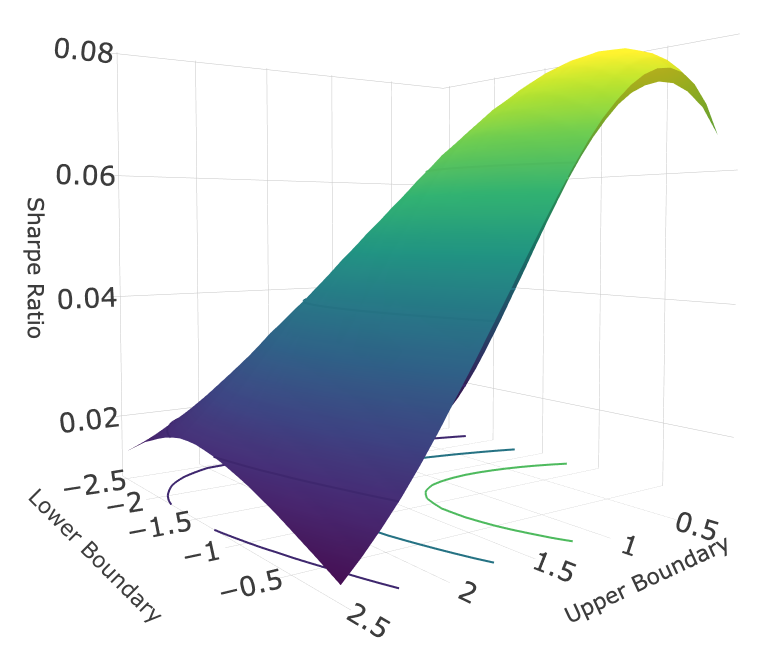}  
  
  \label{fig:M4_strB_sharpe}
\end{subfigure}

%\newline

\begin{subfigure}{.5\textwidth}
  \centering
  \caption{Return of Strategy C, Model 4}
  \includegraphics[scale=0.245]{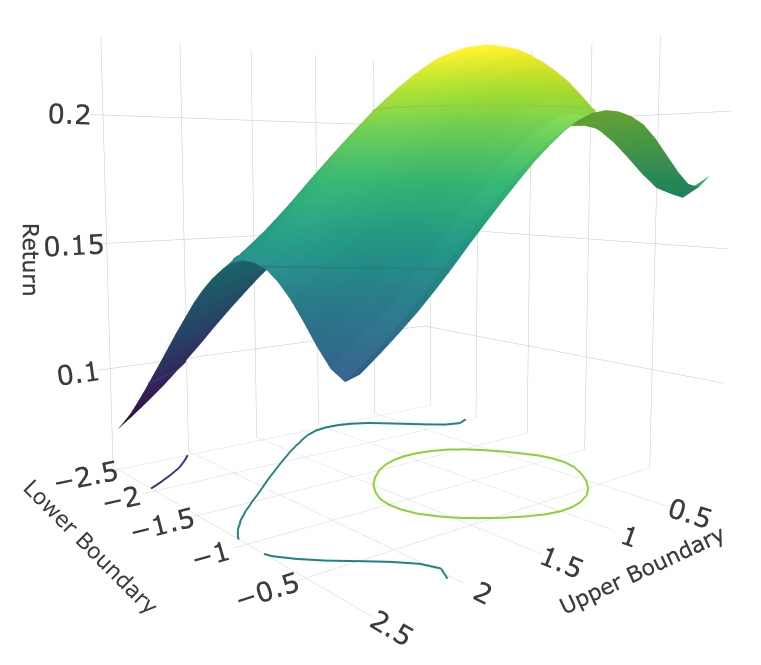}  
  
  \label{fig:M4_strC_tr}
\end{subfigure}
\begin{subfigure}{.5\textwidth}
  \centering
  \caption{Sharpe Ratio of Strategy C, Model 4}
  \includegraphics[scale=0.245]{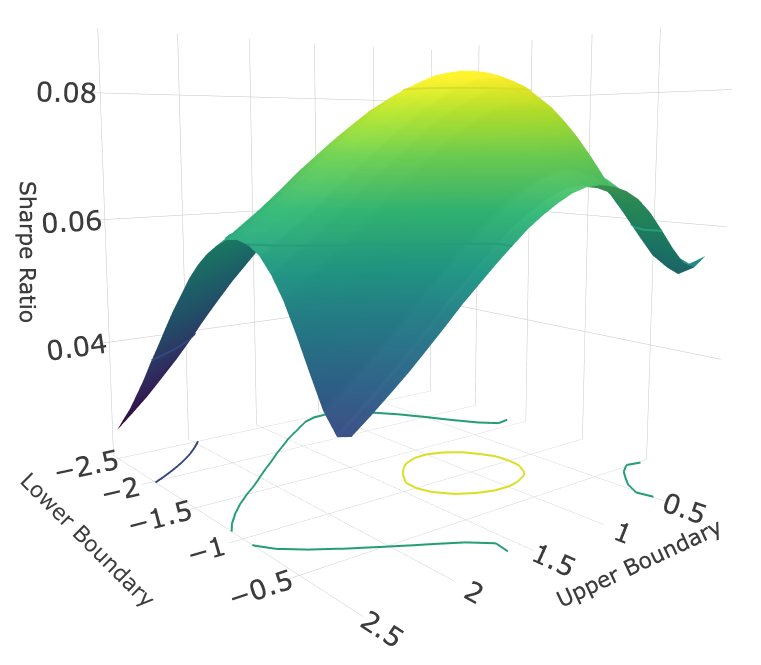}  
  
  \label{fig:M4_strC_sharpe}
\end{subfigure}

\label{fig:model4}
\end{figure}

\clearpage

\begin{sidewaysfigure}
\centering
\caption{Trading signal of Strategy A, B and C on PEP vs KO for Model $\textrm{I}$}
\includegraphics[scale=0.8]{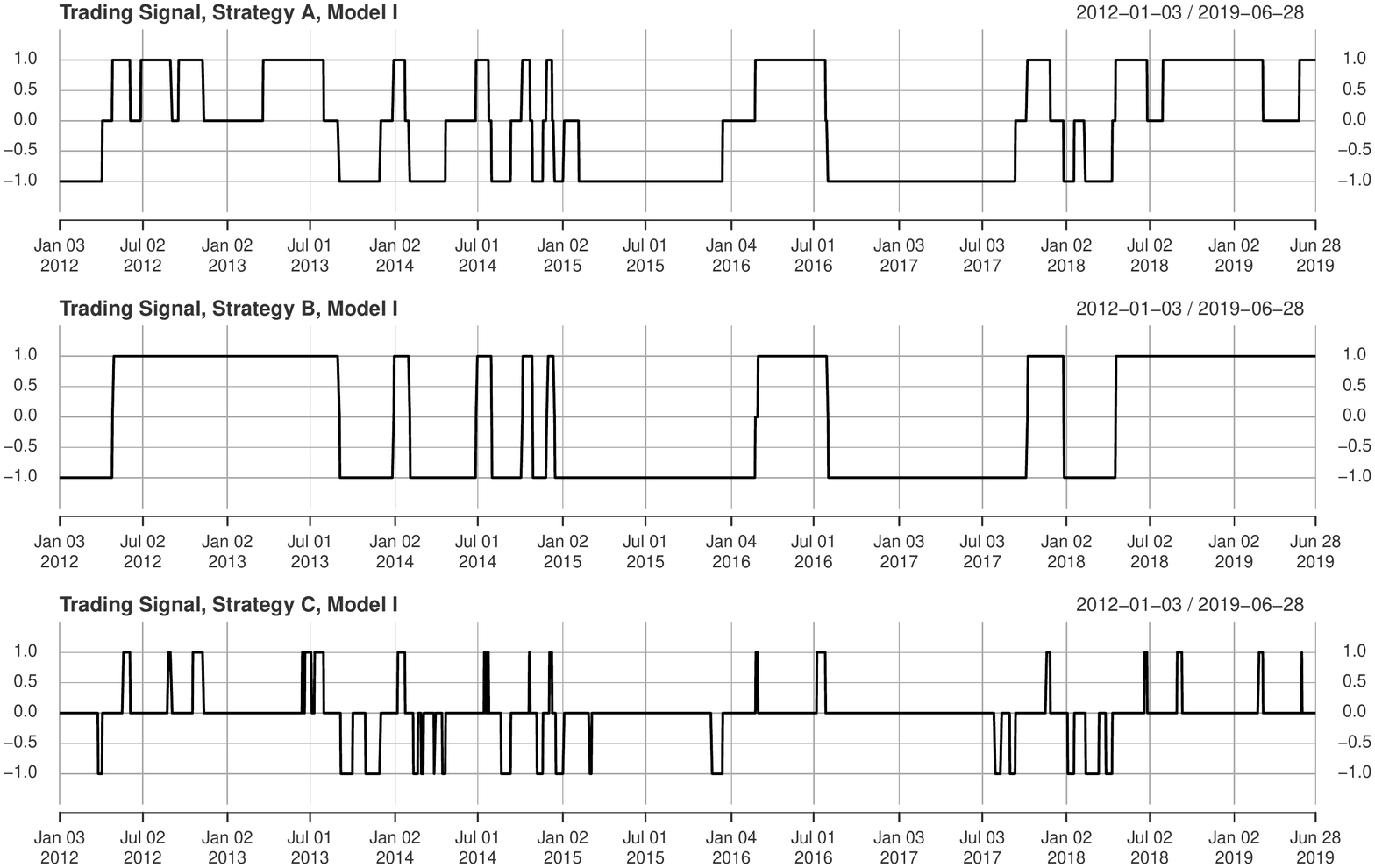}
\captionsetup{justification=raggedright,margin=0.45cm}
\caption*{\footnotesize{Note: When the trading signal is 1 we short PEP and long KO; when the trading signal is -1 we short KO and long PEP; when the trading signal is 0 we clear the position and hold no asset.}}
\label{pepko_signal_m1}
\end{sidewaysfigure}

\begin{sidewaysfigure}
\centering
\caption{Trading signal of Strategy A, B and C on PEP vs KO for Model $\textrm{II}$}
\includegraphics[scale=0.8]{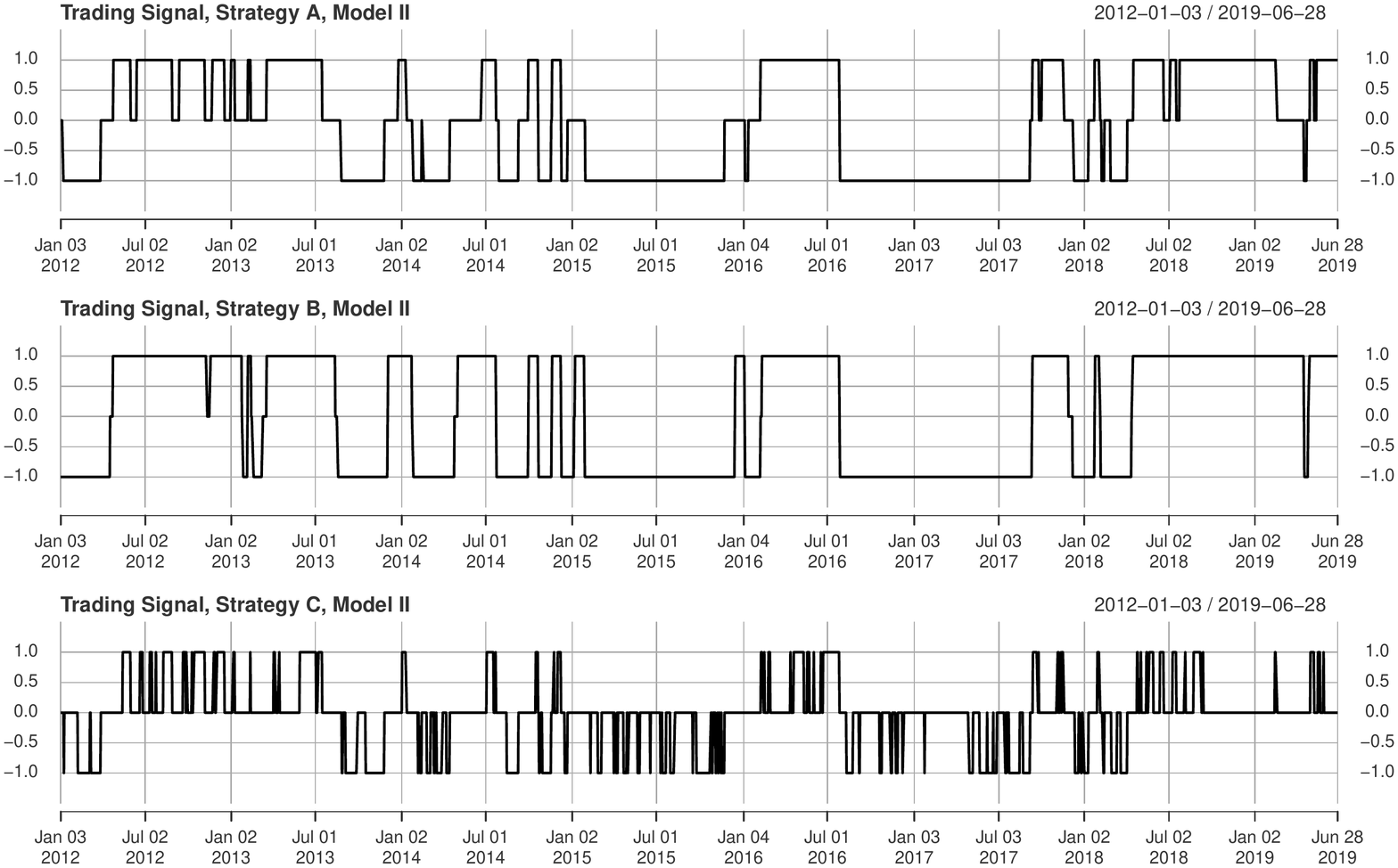}
\captionsetup{justification=raggedright,margin=0.45cm}
\caption*{\footnotesize{Note: When the trading signal is 1 we short PEP and long KO; when the trading signal is -1 we short KO and long PEP; when the trading signal is 0 we clear the position and hold no asset.}}
\label{pepko_signal_m2}
\end{sidewaysfigure}

\begin{sidewaysfigure}
\centering
\caption{Trading Performance of Strategy A, B and C on PEP vs KO for Model $\textrm{I}$}
\includegraphics[scale=0.8]{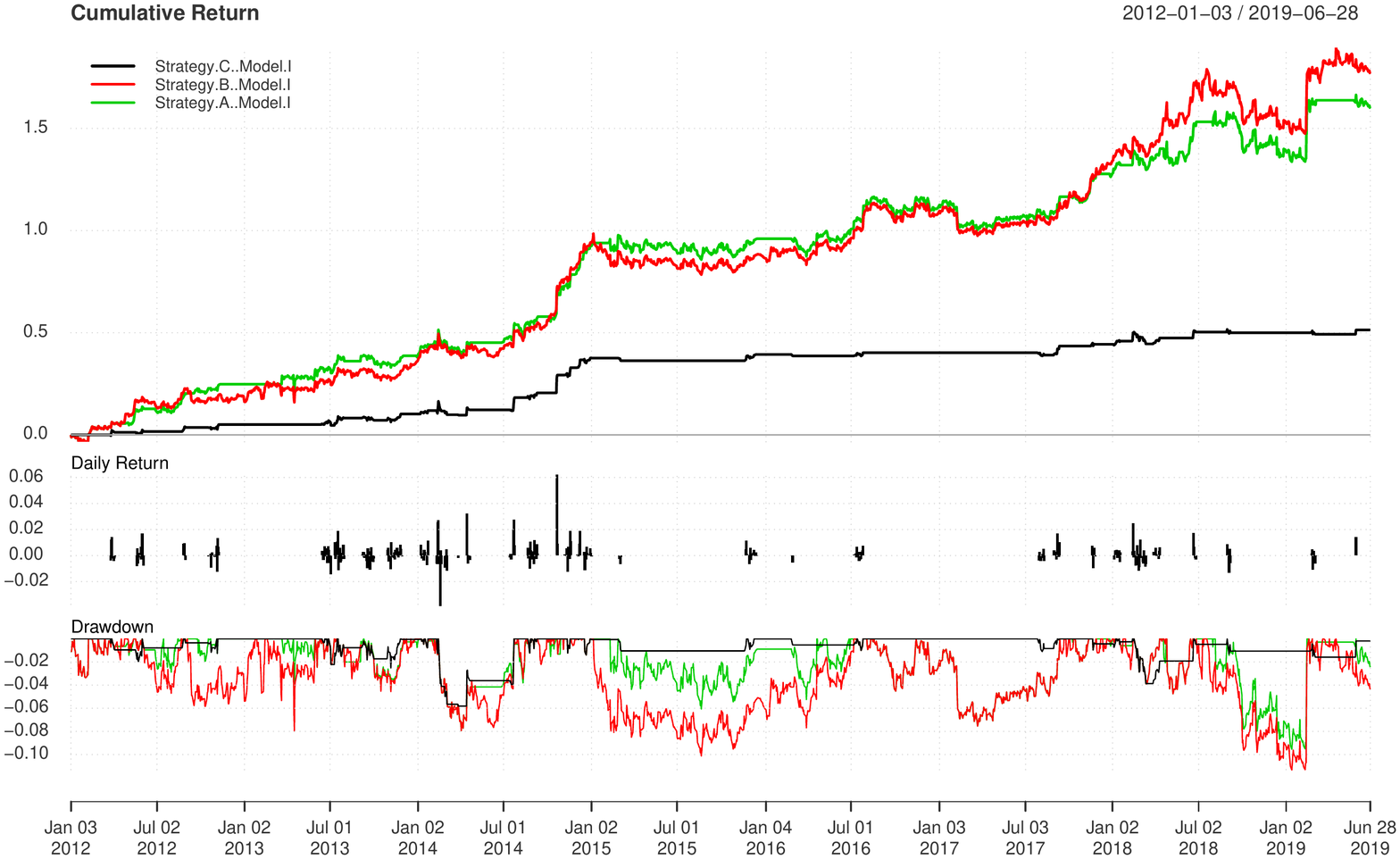}
\captionsetup{justification=raggedright,margin=0.45cm}
\caption*{\footnotesize{Note: Black curves are the results of Stragegy C; red curves are the results of Strategy B; green curves are the results of Strategy A. The Daily Return diagram is only for Strategy C}}
\label{pepko_perf_m1}
\end{sidewaysfigure}

\begin{sidewaysfigure}
\centering
\caption{Trading Performance of Strategy A, B and C on PEP vs KO for Model $\textrm{II}$}
\includegraphics[scale=0.8]{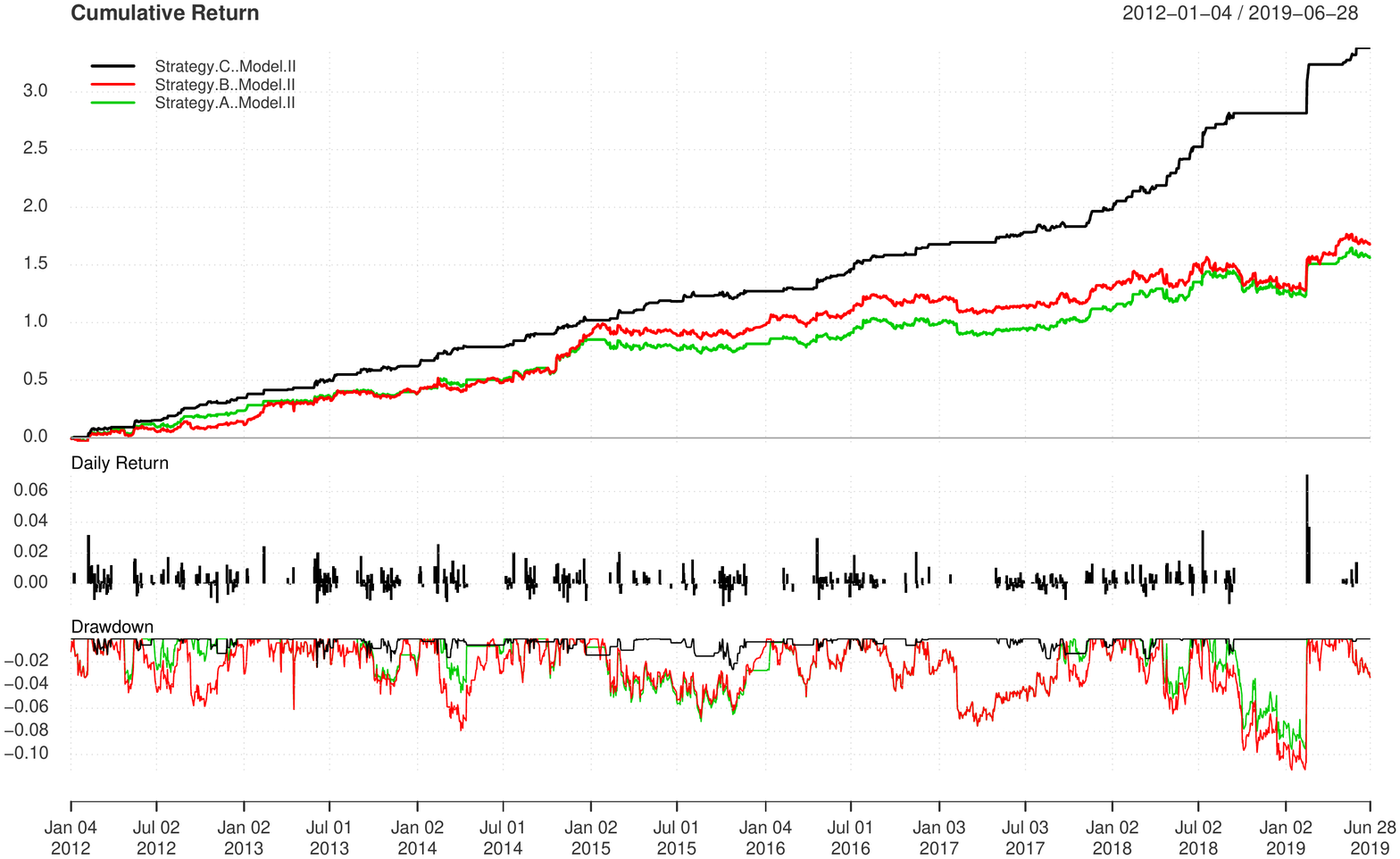}
\captionsetup{justification=raggedright,margin=0.45cm}
\caption*{\footnotesize{Note: Black curves are the results of Stragegy C; red curves are the results of Strategy B; green curves are the results of Strategy A. The Daily Return diagram is only for Strategy C}}
\label{pepko_perf_m2}
\end{sidewaysfigure}

\begin{sidewaysfigure}
\centering
\caption{Trading signal of Strategy A, B and C on EWT vs EWH for Model $\textrm{I}$}
\includegraphics[scale=0.8]{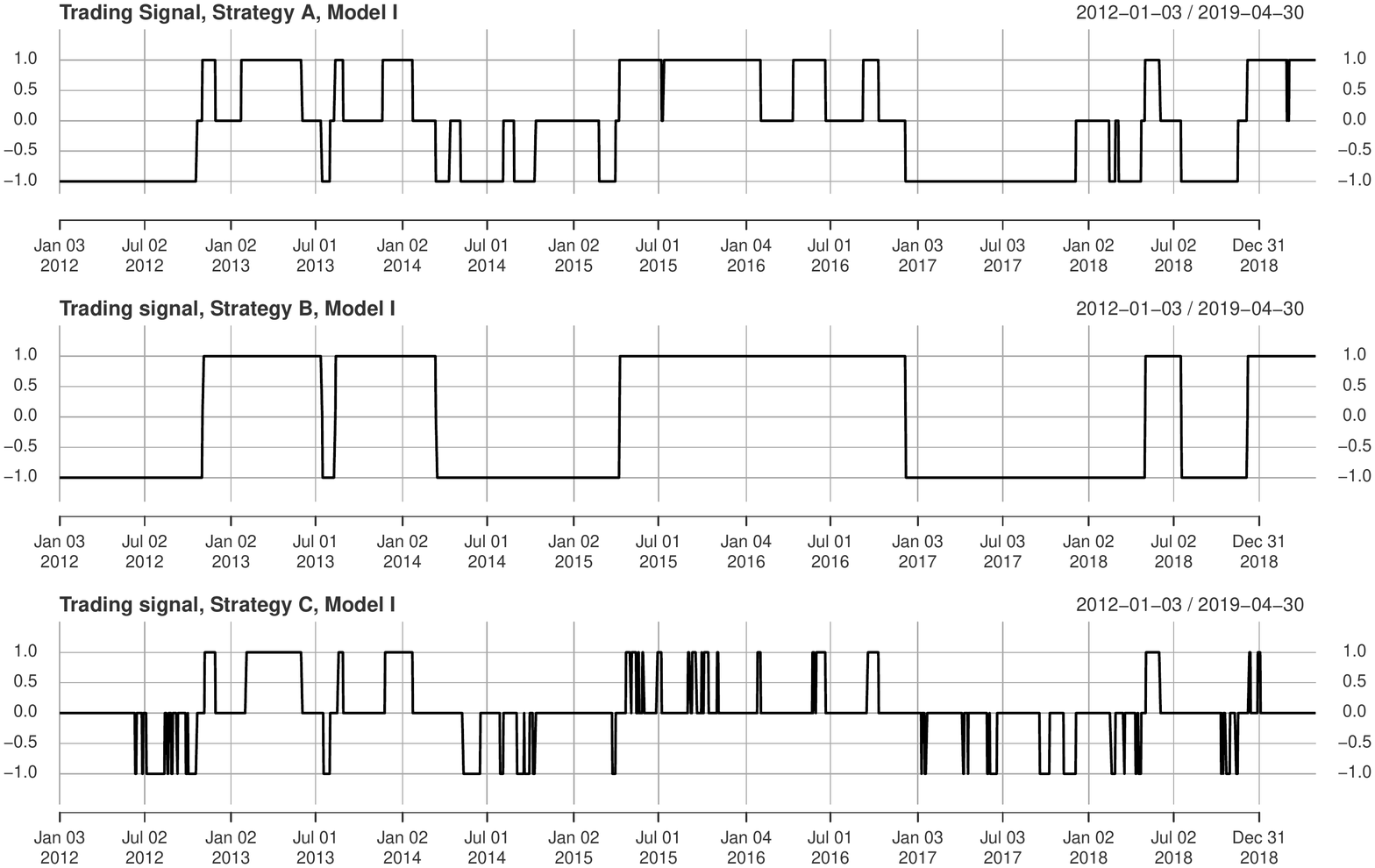}
\captionsetup{justification=raggedright,margin=0.45cm}
\caption*{\footnotesize{Note: When the trading signal is 1 we short EWT and long EWH;when the trading signal is -1 we short EWH and long EWT; when the trading signal is 0 we clear position and hold no asset.}}
\label{ewtewh_signal_m1}
\end{sidewaysfigure}

\begin{sidewaysfigure}
\centering
\caption{Trading signal of Strategy A, B and C on EWT vs EWH for Model $\textrm{II}$}
\includegraphics[scale=0.8]{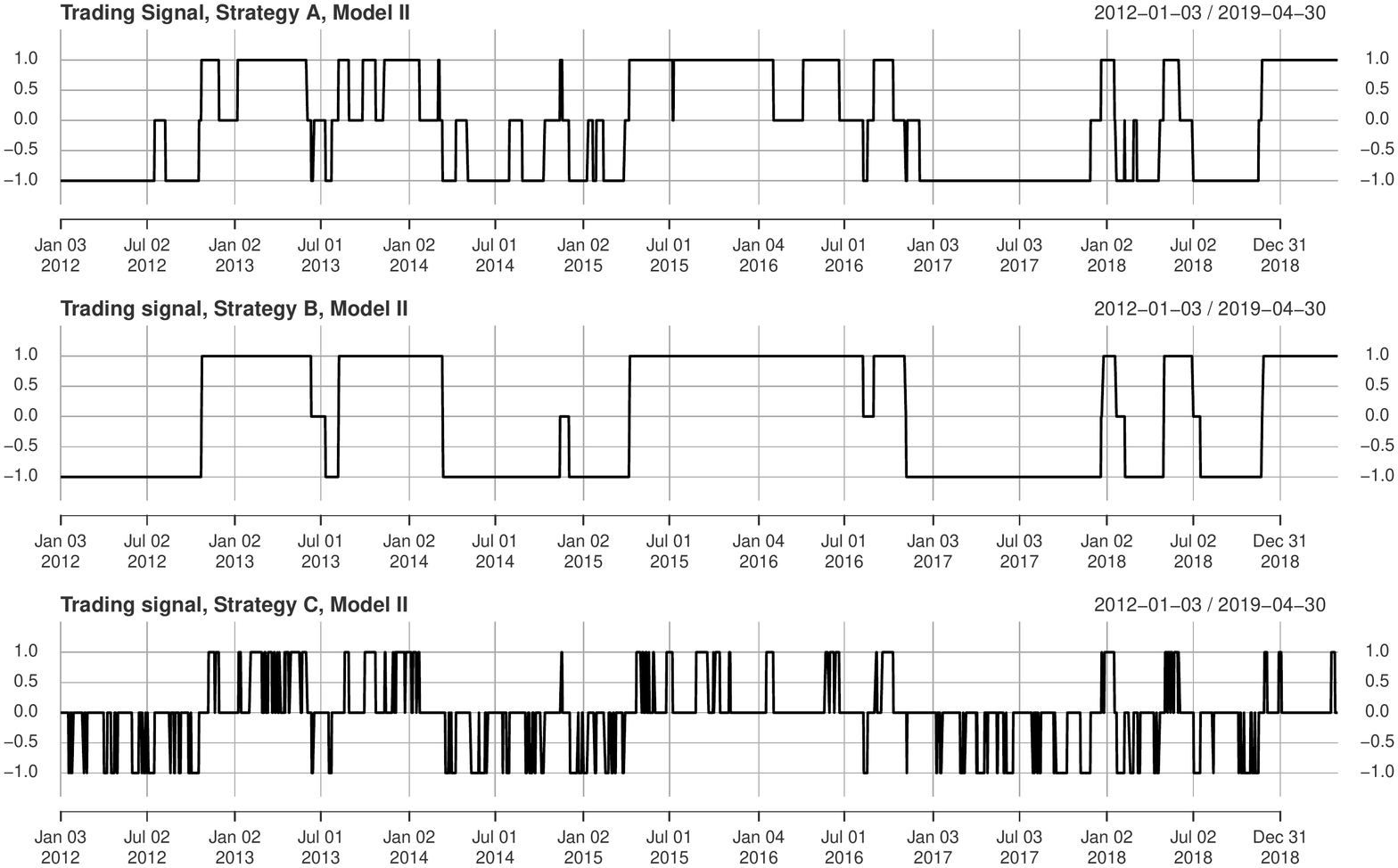}
\captionsetup{justification=raggedright,margin=0.45cm}
\caption*{\footnotesize{Note: When the trading signal is 1 we short EWT and long EWH;when the trading signal is -1 we short EWH and long EWT; when the trading signal is 0 we clear position and hold no asset.}}
\label{ewtewh_signal_m2}
\end{sidewaysfigure}

\begin{sidewaysfigure}
\centering
\caption{Trading Performance of Strategy A, B and C on EWT vs EWH for Model $\textrm{I}$}
\includegraphics[scale=0.8]{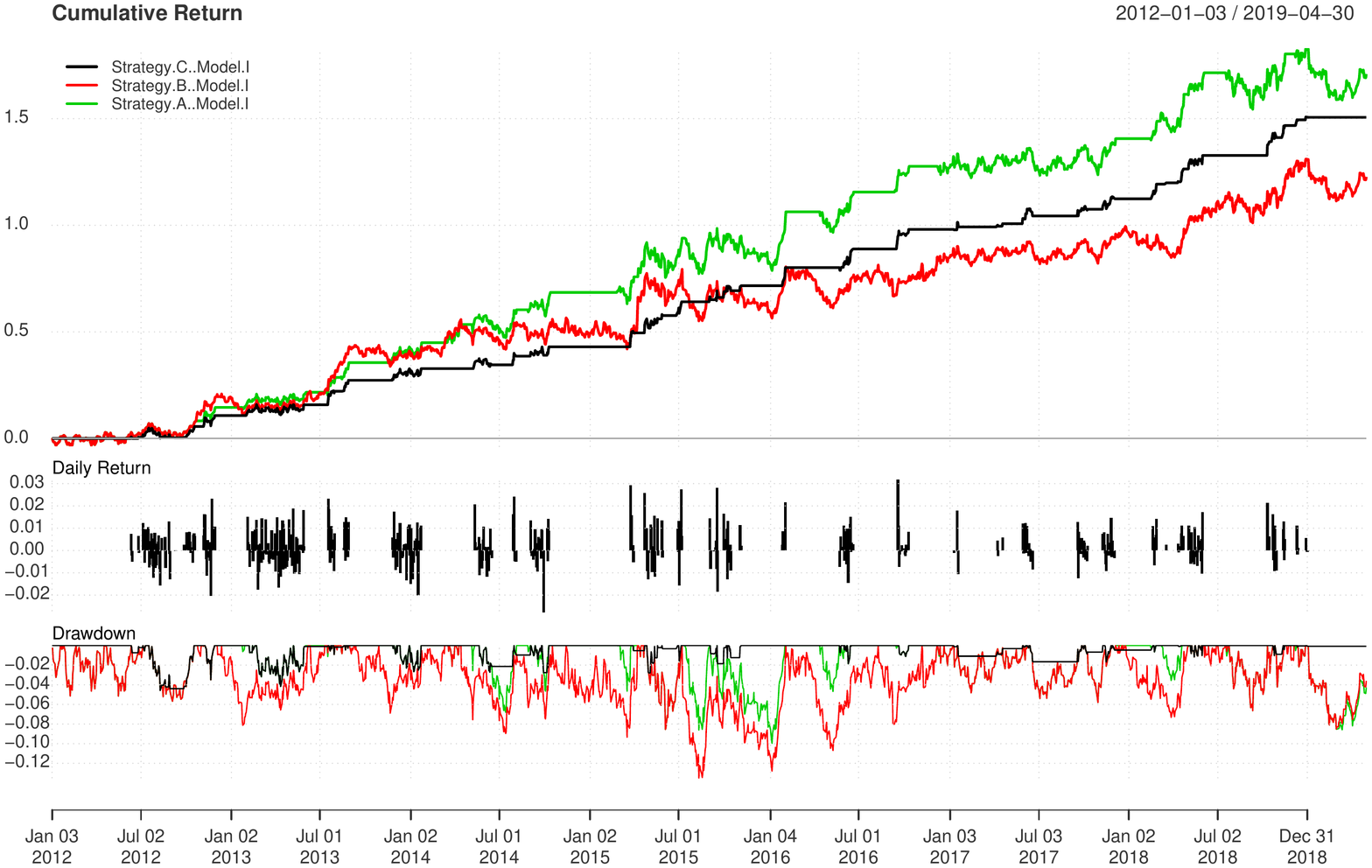}
\captionsetup{justification=raggedright,margin=0.45cm}
\caption*{\footnotesize{Note: Black curves are the results of Stragegy C; red curves are the results of Strategy B; green curves are the results of Strategy A. The Daily Return diagram is only for Strategy C                                                                      }}
\label{ewtewh_perf_m1}
\end{sidewaysfigure}

\begin{sidewaysfigure}
\centering
\caption{Trading Performance of Strategy A, B and C on EWT vs EWH for Model $\textrm{II}$}
\includegraphics[scale=0.8]{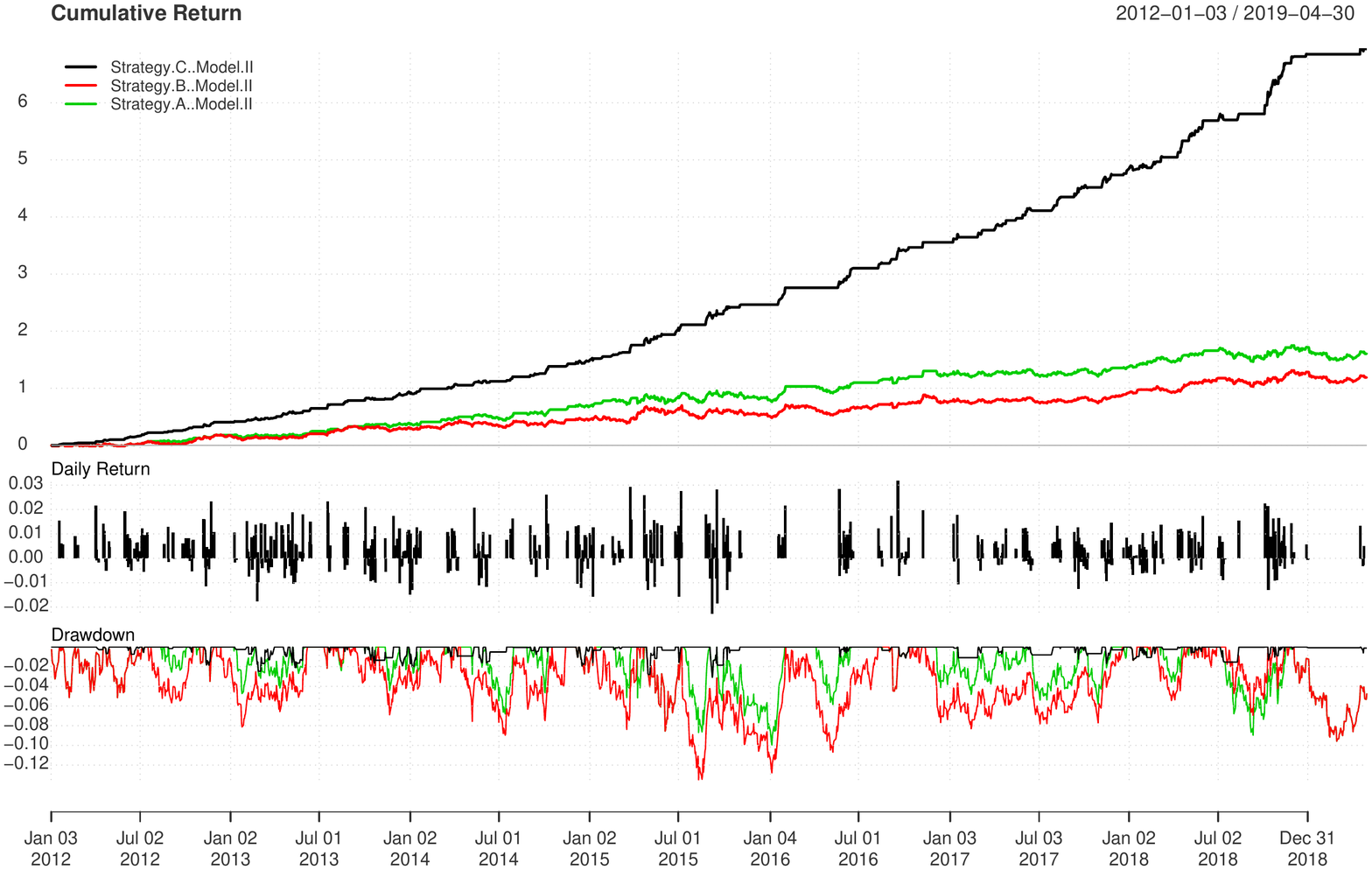}
\captionsetup{justification=raggedright,margin=0.45cm}
\caption*{\footnotesize{Note: Black curves are the results of Stragegy C; red curves are the results of Strategy B; green curves are the results of Strategy A. The Daily Return diagram is only for Strategy C}}
\label{ewtewh_perf_m2}
\end{sidewaysfigure}

\clearpage
\begin{figure}
\centering
\captionsetup{justification=centering}
\caption{Annualized Return and Sharpe Ratio of Pairs Trading on Intergroup Pairs of Large Banks and Small Banks}
\includegraphics[width=15cm,height=8.5cm]{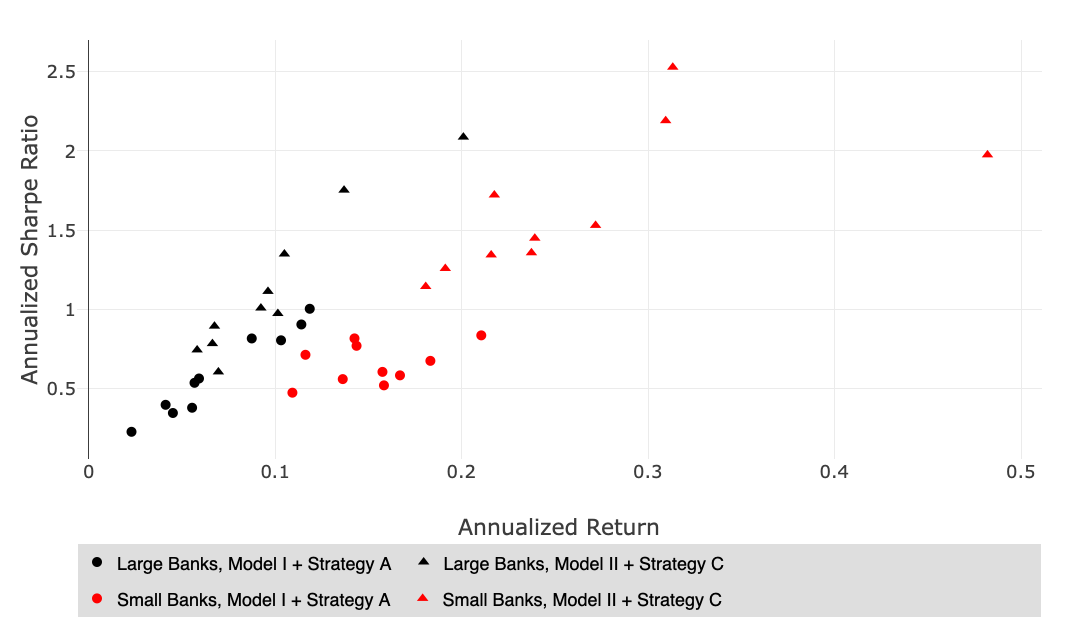}
\captionsetup{justification=raggedright,margin=0.8cm}
\caption*{\footnotesize{Note: Black circles are the performances of Model $\textrm{I}$ + Strategy A on pairs of large banks, red circles are the performances of Model $\textrm{I}$ + Strategy A on pairs of small banks, black triangles are the performances of Model $\textrm{II}$ + Strategy C on pairs of large banks, and red triangles are the performances of Model $\textrm{II}$ + Strategy C on pairs of small banks.}}
\label{fig:inter}
\end{figure}

\begin{figure}
\centering
\caption{Annualized Return and Sharpe Ratio of Pairs Trading on Intragroup Pairs}
\includegraphics[width=15cm,height=8.5cm]{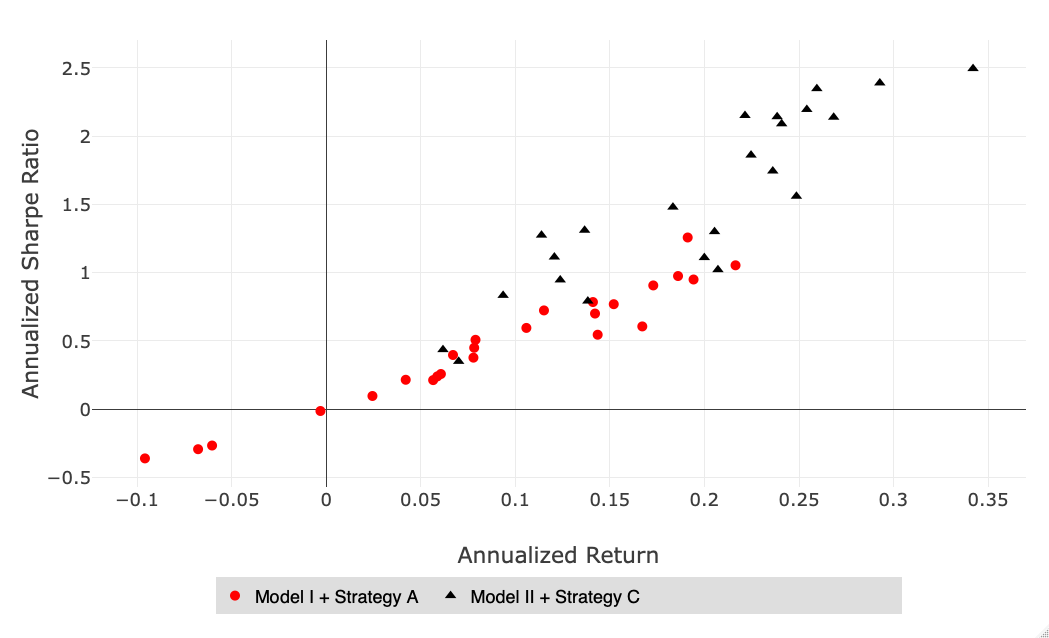}
\captionsetup{margin=0.8cm}
\caption*{\footnotesize{Note: Red circles are the performances of Model $\textrm{I}$ + Strategy A on intragroup pairs: one from the group of large banks and the other one from the group of small banks; the black triangles are the performances of Model $\textrm{II}$ + Strategy C}}
\label{fig:intra}
\end{figure}


\clearpage
\begin{thebibliography}{10}

\bibitem{tourin&yan}Agnès Tourin and Raphael Yan, 2013, 
\textit{Dynamic pairs trading using the stochastic control approach}, Journal of Economic Dynamics and Control, 37 (2013) 1972-1981.

\bibitem{sahalia}Ait-Sahalia, Y., 1996, 
\textit{Testing Continuous-Time Models
of the Spot Interest Rate}, Review of Financial Studies, 9, 385-426

\bibitem{avelaneda&lee}Avellaneda, M., and J.-H. Lee. 2010. 
\textit{Statistical arbitrage in the US equities market}. Quantitative Finance 10:761--782.

\bibitem{mandelbrot71}Benoit B. Mandelbrot, 1971, 
\textit{When Can Price be Arbitraged Efficiently? A Limit to the Validity of the Random Walk and Martingale Models}, The Review of Economics and Statistics, Vol. 53, No. 3 (Aug., 1971), pp. 225-236

\bibitem{bogomolov}Bogomolov, T. 2013. 
\textit{Pairs trading based on statistical variability of the spread process}, Quantitative Finance 13:1411--1430.

\bibitem{carlos&adrian&jorge}Carlos Eduardo de Moura, Adrian Pizzinga and Jorge Zubelli (2016),
\textit{A pairs trading strategy based on linear state space models and the Kalman filter}, Quantitative Finance

\bibitem{clegg&matthew&krauss}Clegg, Matthew and Krauss, Christopher. 2018, 
\textit{Pairs trading with partial cointegration}, Quantitative Finance 18 (1), 121--138.

\bibitem{cummins&bucca}Cummins, Mark and Bucca, Andrea, 2012, 
\textit{Quantitative spread trading on crude oil and refined products markets}, Quantitative Finance. Dec2012, Vol. 12 Issue 12, p1857-1875.

\bibitem{david}David A. Hsieh, 1989, 
\textit{Testing for Nonlinear Dependence in Daily Foreign Exchange Rates}, The Journal of Business, Vol. 62, No. 3 (Jul., 1989), pp. 339-368

\bibitem{ding&granger}Ding, Z., Granger, C.W.J. 
\textit{Modeling volatility persistence of speculative returns: A new approach}, Journal of Econometrics, 1996, vol. 73, issue 1, 185-215

\bibitem{fama&macbeth}E. F. Fama and James D. MacBeth. 
\textit{Risk, return, and equilibrium}, The Journal of Political Economy 791.1 (1971), pp. 30--55

\bibitem{elliott&vanderhoek}Elliott, R. J., J. Van Der Hoek, and W. P. Malcolm. 2005. 
\textit{Pairs trading}. Quantitative Finance 5:271--276

\bibitem{elliott&bradrania}Elliott, R. J. ; Bradrania, R, 2018, 
\textit{Estimating a regime switching pairs trading model}, Quantitative Finance, 2018, Vol.18(5), pp.877-883

\bibitem{fama}Eugene F. Fama, 1970, 
\textit{Efficient Capital Markets: A Review of Theory and Empirical Work}, The Journal of Finance, Vol. 25, No. 2, Papers and Proceedings of the TwentyEighth Annual Meeting of the American Finance Association New York, N.Y. December, 28-30, 1969 (May, 1970), pp. 383-417

\bibitem{gatev&goetzmann&rouwenhorst}Gatev, E.G., Goetzmann, W.N. and Rouwenhorst, K.G. (2006). 
\textit{Pairs Trading: Performance of a Relative Value Arbitrage Rule}. The Review of Financial Studies, 19, 797-827.

\bibitem{scheinkman&lebaron}Jose A. Scheinkman and Blake LeBaron, 1989, 
\textit{Nonlinear dynamics and stock returns, The Journal of Business}, Vol. 62, No. 3 (Jul., 1989), pp. 311-337

\bibitem{suzuki}Kiyoshi Suzuki, 2018, 
\textit{Optimal pair-trading strategy
over long/short/square positions---empirical study}, Quantitative Finance, Volume 18, 2018 - Issue 1

\bibitem{kon}Kon S, 1984, 
\textit{Models of stock returns: a comparison}, J. Finance XXXIX 147--65

\bibitem{mandelbrot63}Mandelbrot B, 1963,
\textit{The variation of certain speculative prices}, J. Business XXXVI 392--417

\bibitem{barndorff&shephard}Ole E. Barndorff-Nielsen, and Neil Shephard, 2001,
\textit{Non-Gaussian Ornstein-Uhlenbeck-based models and some of their uses in financial economics}, J. R. Statist. Soc. B (2001) 63, Part 2, pp. 167-241

\bibitem{rad&low&faff}Rad, H., R. K. Y. Low, and R. Faff. 2016. 
\textit{The profitability of pairs trading strategies: distance, cointegration and copula methods}. Quantitative Finance 16:1541--1558.

\bibitem{rama}Rama Cont, 2001, 
\textit{Empirical properties of asset returns: stylized facts and statistical issues}, Quantitative Finance VOL 1 (2001) 223--236

\bibitem{sergio&frank&ivan}Sergio M. Focardi, Frank J. Fabozzic, Ivan K. Mitov, 2016, 
\textit{A new approach to statistical arbitrage: Strategies based on dynamic factor models of prices and their performance}. Journal of Banking \& Finance, 65 (2016) 134-155

\bibitem{stubinger&endres}Stübinger, Johannes and Endres, Sylvia, 2018, 
\textit{Pairs trading with a mean-reverting jump-diffusion model on high-frequency data}, Quantitative Finance Volume 18, 2018 - Issue 10

\bibitem{bollerslev&chou&kroner}Tim Bollerslev, Ray Y. Chou and Kenneth F. Kroner, 1992,
\textit{ARCH modeling in finance: A review of the theory and empirical evidence}, Journal of Econometrics 52 (1992) 5-59.

\bibitem{vidyamurthy}Vidyamurthy, G., 2004. 
\textit{Pairs trading: Quantitative methods and analysis}. J. Wiley, Hoboken, N.J.

\bibitem{bai&wu}Yang Bai and Lan Wu, 2018, 
\textit{Analytic value function for optimal regime-switching pairs trading rules}, Quantitative Finance, 2018, Vol.18(4)

\bibitem{yang&qiao&beling&scherer}Yang, S. Y., Qiao, Q., Beling, P. A., Scherer, W. T., Kirilenko, A. A., 2015. 
\textit{Gaussian process-based algorithmic trading strategy identification}. Quantitative Finance 0 (0), 1--21.

\bibitem{lei&xu}Yaoting Lei and Jing Xu, 2015, 
\textit{Costly arbitrage through pairs trading}, Journal of Economic Dynamics \& Control, 56 (2015) 1-19

\bibitem{zeng&lee}Zeng, Z., Lee, C. G., 2014. 
\textit{Pairs trading: optimal thresholds and profitability}. Quantitative Finance 14 (11), 1881--1893.

\bibitem{ding&granger&engle}Zhuanxin Ding, Clive W.J. Granger, Robert F. Engle (1993) 
\textit{A long memory property of stock market returns and a new model},Journal of Empirical Finance, Volume 1, Issue 1, 1993, Pages 83-106

\end{thebibliography}
\end{document}